\documentclass[preprint,amsmath,amssymb,showpacs]{revtex4}
\usepackage{epsfig}
\usepackage{dcolumn}
\usepackage{graphicx}
\usepackage{subfigure}

\begin{document}

\title{Relativistic First-Principles Full Potential Calculations  
of Electronic and Structural Properties 
of group IIIA-VA  semiconductors based on Zeroth Order Regular Approximation (ZORA) Hamiltonian.}

\author{Eugene S. Kadantsev}
\affiliation{Quantum Theory Group, Institute for Microstructural Sciences,
National Research Council, Ottawa, Canada K1A 0R6}
\email{ekadants@babylon.phy.nrc.ca}
\date{\today}

\begin{abstract}
First-principles full potential calculations based on  Zeroth Order Regular 
Approximation (ZORA) relativistic Hamiltonian and  Kohn-Sham form of Density 
Functional Theory (KS DFT) in local spin density approximation (LSDA) 
are reported for group 
IIIA-VA ({\rm InAs}, {\rm GaAs}, {\rm InP})  semiconductors. 
The effects of relativity are elucidated by performing fully 
relativistic, scalar relativistic, and nonrelativistic 
calculations. Structural and electronic band structure parameters 
are determined including split-off energies, band gaps,  
and deformation potentials. The nature of 
chemical bonding at the equilibrium and under hydrostatic 
strain is investigated using projected  (PDOS) and overlap population 
weighted density of states (OPWDOS). ZORA results are compared with 
Augmented Plane Wave plus Local Orbitals method (APW+lo), and experiment. Viability and 
robustness of the ZORA relativistic Hamiltonian for investigation of electronic and 
structural properties of semiconductors is established. 
\end{abstract}
\maketitle

\section{Introduction} 

There is a great interest in electronic and structural 
properties of group IIIA-VA  materials
due to their wide spread applications in semiconductor devices. 
In particular, {\rm InAs/GaAs} and {\rm InAs/InP} semiconductor quantum 
dots (QDs)~\cite{Hawrylak1997} 
have shown great promise~\cite{Dalacu2009} in quantum information applications 
such as is the generation of entangled photon pairs (EPPs) on 
demand~\cite{Shields2007, Stevenson2006, Akopian2006, Greilich2006}. 

Atomistic modeling of semiconductor nanostructures may 
require input from accurate Density Functional~\cite{Hohenberg1964,Kohn1965,vonBarth1972} calculations 
in cases when experimental data is not available.
Therefore, it is important to understand which material parameters are well 
reproduced with ``standard DFT" 
and this work is a contribution in this area. Three factors determine
accuracy of Density Functional calculations 1) Model exchange-correlation functional; 
2) Representation of single-particle orbitals 
(atomic orbitals, plane-waves, real space grids) and 
representation of ion-electron interaction ({\it ab initio}  pseudopotentials, full potential schemes);
3) Treatment of relativity. The assessment of 
the accuracy of exchange-correlation functionals is beyond the scope of  this contribution. 
The main objective of this work is to perform a detailed study of structural and 
electronic structure properties of {\rm InAs}, {\rm InP}, and {\rm GaAs} semiconductors 
using highly accurate representation of single-particle orbitals and ion-electron interaction
and to assess the role of relativity in these calculations.

First-principles calculations on group IIIA-VA 
semiconductors based on Kohn-Sham form of Density Functional 
Theory~\cite{Hohenberg1964,Kohn1965,vonBarth1972} have already been 
performed in the past~\cite{VandeWalle1989,RevModPhys.75.863,wei1998,wei1999,wei2006,wei2009}. 
The computational approach 
used in these calculations generally evolved from {\it ab initio} pseudopotential
calculations to more elaborate full potential (FP) augmentation 
schemes~\cite{Singh1994, Schwarz2001, Sjostedt2000, Madsen2001}  
such as Linearized Augmented Plane Wave (LAPW) and 
Augmented Plane Wave plus local orbitals (APW+lo)
methods. LAPW and APW+lo have now become the methods of choice 
when accuracy considerations have the highest priority. 

There is also a less known full potential method due to te Velde and 
Baerends~\cite{BAND0,Velde1991} which makes use of Bloch basis set made up 
of numerical and Slater type atomic orbitals (NAO/STO basis), a basis which is excellent for the accurate 
representation of the electron density. The approach due to te Velde and Baerends 
is implemented in {\rm BAND} 
program~\cite{BAND} and is capable of treating chemical elements throughout periodic 
table using Zeroth Order Regular Approximation (ZORA) relativistic 
Hamiltonian~\cite{vanLenthe1993,vanLenthe1994, vanLenthe1996A,vanLenthe1996B,vanLenthe1999}. 
ZORA  approach is designed to capture scalar relativistic 
effects such as $s$  and $p$ ``orbital contraction" (stabilization)
and  $d$ ``orbital expansion" (destabilization) as well 
as spin-orbital splitting for electrons with angular momentum $l > 0$. 
While ZORA Hamiltonian is very well established among quantum chemists,
relativistic ZORA calculations on solids are 
much less common~\cite{Philipsen1997,Kootstra2000,Olsen2003,Romaniello2005,Romaniello2007,
Kadantsev2009G}, especially, in comparison with a large volume of calculations employing LAPW and APW+lo
methods. Therefore, further assessment of ZORA performance in solids 
is important.

In this work, I  will employ ZORA Hamiltonian to perform a detailed
study of structural and electronic band structure 
properties of {\rm InAs, InP} , and {\rm GaAs}
semiconductors in zinc-blende phase. The effects of relativity are 
elucidated  by performing three sets 
of calculations  1) nonrelativistic 2) scalar relativistic and
3) relativistic with variational treatment of 
spin-orbital coupling (or fully relativistic). ZORA Hamiltonian is applied to 
calculate electronic band structures, band gaps, and deformation 
potentials. The results obtained with ZORA Hamiltonian are compared to 
those obtained with APW+lo method, and with experiment. 
 
I perform a detailed analysis of electronic structure of 
{\rm InAs, InP} and {\rm GaAs}
in terms of Projected and Overlap Population Weighted Density of States 
(PDOS/OPWDOS analysis). Whereas it is a relatively general practice 
to report and discuss PDOS, the usage OPWDOS is  much less common. 
The OPWDOS analysis, popularized~\cite{Hoffmann1988} by the Nobel Prize 
winner Roald Hoffmann, provides a clear pictorial representation  of bonding, non-bonding, 
and anti-bonding orbital interactions and is deemed useful. 
I will demonstrate the usage of OPWDOS 
plots and explain how they add to our understanding of chemical bond 
in {\rm InAs, InP}, and {\rm GaAs}. 

Atomic units $\hbar = e = m_e = 1$ are used throughout unless otherwise specified.

\section{Computational Approach}

The first-principles KS DFT calculations with ZORA Hamiltonian are carried out  
with~{\rm BAND} program~\cite{BAND0,Velde1991,BAND}. {\rm BAND}  makes use of 
periodic boundary conditions (PBC) and explicit Bloch basis composed of 
numerical and Slater type atomic orbitals (NAO/STAO basis). I also perform calculations 
with APW+lo method~\cite{Sjostedt2000, Madsen2001} as implemented in {\rm EXCITING} 
program~\cite{EXCITING}. Both methods are capable of including spin-orbital 
coupling variationally and I choose to do so. To clarify effects 
of relativity, I also present results of scalar relativistic and 
nonrelativistic calculations. 

The relativistic effects in {\rm BAND} are treated  
in the Zeroth Order Regular Approximation (ZORA) approach of van Lenthe and 
co-workers~\cite{vanLenthe1993,vanLenthe1994, vanLenthe1996A,vanLenthe1996B,vanLenthe1999}. 
The details of ZORA implementation in {\rm BAND} are given in Ref.~\cite{Philipsen1997}. 
In ZORA approach, the kinetic energy operator is replaced by the 
ZORA expression $\hat{T}^{ZORA}$
\begin{equation} 
-\frac{1}{2} \mathbf{p}^2 \Rightarrow \hat{T}^{ZORA} = \vec{\sigma} \cdot \mathbf{p} 
\frac{c^2}{2 c^2 - V_{SAPA}(\mathbf{r})} \vec{\sigma} \cdot \mathbf{p}, 
\end{equation} 
where $\mathbf{p}$ is the ``momentum operator" ($\mathbf{p} = -i \nabla$ in 
the absence of a magnetic field); $V_{SAPA}(\mathbf{r})$ -- is a sum of atomic 
potentials (SAPA), an approximation to the total effective potential in the ZORA kinetic 
energy operator; and $\vec{\sigma} = 
\{\sigma_x, \sigma_y, \sigma_z \}$ -- is a vector made up of the Pauli matrices. 
Introducing the notation 
\begin{equation} 
K = \frac{1}{1-V_{SAPA}(\mathbf{r})/2c^2}, 
\end{equation} 
the ZORA kinetic energy becomes 
\begin{equation} 
\hat{T}^{ZORA} = \vec{\sigma} \cdot \mathbf{p} 
\frac{K}{2} \vec{\sigma} \cdot \mathbf{p} = \mathbf{p} \frac{K}{2} \mathbf{p} + \frac{1}{2} \vec{\sigma} \cdot 
\left ( \nabla K \times \mathbf{p} \right ). 
\end{equation} 

In the last equation, $\hat{T}^{ZORA}$ was split into the so-called scalar relativistic 
$\hat{T}^{ZORA}_{SR}$ and spin-orbital  $\hat{T}^{ZORA}_{SO}$ terms, where 
$\hat{T}^{ZORA}_{SO}= 1/2 \vec{\sigma} \cdot  ( \nabla K \times \mathbf{p}  )$.  
The nonrelativistic limit can be obtained by setting $K \rightarrow 1$. 
I will refer to calculations with $\hat{T}^{ZORA}$, $\hat{T}^{ZORA}_{SR}$,
and $K = 1$ as (fully) relativistic (ZORA FREL), scalar relativistic (ZORA SREL), and 
nonrelativistic (NREL), respectively.  

In my {\rm BAND} calculations, I use  basis set of triple zeta quality 
({\rm TZ2P} in {\rm BAND}'s notation) taken from the program's database.
The core states are obtained from the full potential atomistic calculations 
and are kept frozen during the self consistent 
field (SCF) procedure. The valence states are expanded in terms of the NAO/STAO Bloch basis 
set functions orthogonalized on the core states (VOC basis). 
The Hamiltonian matrix elements are 
evaluated using highly accurate numerical integration scheme~\cite{Boerrigter1988}.
The Brillouin zone integration is carried out using  accurate quadratic tetrahedron 
method~\cite{Wiesenekker1991,Wiesenekker1988} with 65 symmetry unique $\mathbf{k}$-points spanning 
the irreducible Brillouin zone (IBZ). The ``default" convergence criteria are used to 
terminate the SCF procedure. 

APW+lo calculations are carried out with {\rm EXCITING} 
program~\cite{EXCITING}. The local orbitals and linearization energies 
are taken from the program's database.  The ``core" states  are treated fully relativistically 
and self-consistently in the spherical approximation, whereas the ``valence" states are treated using 
the second-variational Hamiltonian. The IBZ is sampled using $\{8 \times 8 \times 8\}$ uniform 
mesh of $\mathbf{k}$-points.   

The exchange-correlation is treated within the local spin density approximation 
(LSDA)~\cite{Kohn1965,vonBarth1972}. {\rm BAND}  and APW+lo LSDA calculations 
are carried out using Vosko-Wilk-Nusair~\cite{Vosko1980} and Perdew-Wang~\cite{Perdew1992} 
parameterization of the correlation energy, respectively. 
The calculations are performed using ``primitive" face-centered cubic cell with two atoms 
per cell. The zinc-blende crystal structure was assumed. 

The structural parameters are obtained by varying the 
lattice constant (from -20 to 20\% of the equilibrium volume) and fitting the 
total energies to the Murnaghan equation of state~\cite{Murnaghan1937}. 

The electronic structure is analyzed 
using projected density of states (PDOS) and overlap population weighted 
density of states (OPWDOS). PDOS is defined for a function $\chi_{\mu}$
or a set of functions $\{\chi_{\mu}\}$ and has large values at energies where $\chi_{\mu}$ 
(or $\{\chi_{\mu}\}$) has large contributions in eigenstates (bands, 
molecular orbitals). PDOS can  also be large in energy intervals with a large number of states.
The weights in my PDOS are derived from Mulliken population 
analysis~\cite{Mulliken1955,Bickelhaupt1996}. The $i$th eigenstate of the Kohn-Sham 
Hamiltonian $\psi_{i \mathbf{k}}$ is expanded in a finite basis 
\begin{equation}
\psi_{i \mathbf{k}}(\mathbf{r}) = \sum_{\mu} c_{\mu i} (\mathbf{k}) \chi_{\mu \mathbf{k}}(\mathbf{r}), 
\end{equation}
where $c_{\mu i}(\mathbf{k})$ are expansion coefficients and $\chi_{\mu \mathbf{k}}$ are 
basis set functions -- Bloch sums of equivalent atomic orbitals. The gross population 
of $\chi_{\mu \mathbf{k}}$ for the eigenstate 
$\psi_{i \mathbf{k}}$ is
\begin{equation}
GP_{i \mathbf{k}}(\chi_{\mu \mathbf{k}}) = 
\frac{1}{2} \sum_{\nu} \left( c_{\mu i}(\mathbf{k}) c_{\nu i}^{\dagger}(\mathbf{k}) 
S_{\mu \nu}(\mathbf{k})+c_{\mu i}^{\dagger}(\mathbf{k}) c_{\nu i}(\mathbf{k}) S_{\nu \mu}(\mathbf{k}) \right) 
\end{equation} 
and PDOS for function $\chi_{\mu}$  is 
\begin{equation}
PDOS_{\chi_{\mu}}(E) = \sum_i \sum_{\mathbf{k}} GP_{i \mathbf{k}} (\chi_{\mu \mathbf{k}}) L(E-E_{i\mathbf{k}}), 
\end{equation}
where $E_{i\mathbf{k}}$ is Kohn-Sham eigenvalue corresponding to eigenstate $\psi_{i \mathbf{k}}$ and $L$ is 
a Lorentzian broadening function.

OPWDOS is defined for two functions $\chi_{\mu}$ and $\chi_{\nu}$ or 
between two sets of functions ($\{\chi_{\mu}\}$ and $\{\chi_{\nu}\}$) and has large positive 
or negative values depending on whether the 
interaction between $\chi_{\mu}$ and $\chi_{\nu}$ ($\{\chi_{\mu}\}$ and $\{\chi_{\nu}\}$) is bonding 
or anti-bonding, respectively.  The use of these plots is demonstrated in Ref.~\cite{Hoek1989}.  
The overlap population of two orbitals and OPWDOS are defined as
\begin{eqnarray}
&& OP_{i \mathbf{k}}(\chi_{\mu \mathbf{k}},\chi_{\nu \mathbf{k}}) = 
 c_{\mu i}(\mathbf{k}) c_{\nu i}^{\dagger}(\mathbf{k}) 
S_{\mu \nu}(\mathbf{k})+c_{\mu i}^{\dagger}(\mathbf{k}) c_{\nu i}(\mathbf{k}) S_{\nu \mu}(\mathbf{k}), \nonumber\\
&& OPWDOS_{\chi_{\mu}, \chi_{\nu}}(E)  = \sum_i \sum_{\mathbf{k}} 
OP_{i \mathbf{k}}(\chi_{\mu \mathbf{k}},\chi_{\nu \mathbf{k}}) L(E-E_{i\mathbf{k}}) .
\end{eqnarray}

\section{Results}

\subsection{{\rm InP}, {\rm GaAs}, and {\rm InAs}: Structural Parameters}

\begin{center}
\begin{figure*}
\caption{\label{FIG:IIIVA} ``Spherical atom" energy levels.}
{\epsfig{file=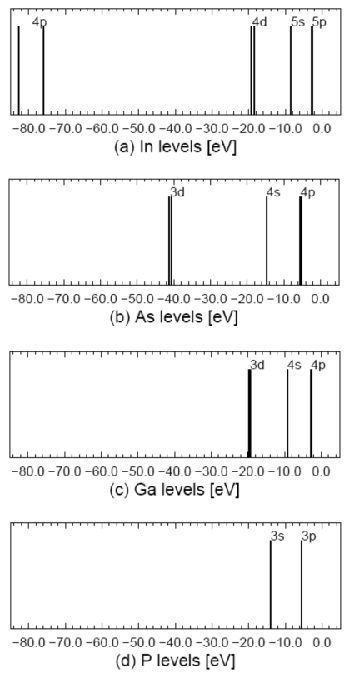,
width=0.70\textwidth,keepaspectratio,angle=0}}
\end{figure*}
\end{center}

Fig.~\ref{FIG:IIIVA} shows energy level diagram for ``spherically symmetric" 
{\rm In}, {\rm As}, {\rm Ga}, and {\rm P} atoms. Fig.~\ref{FIG:IIIVA} shows 
that group IIIA atomic species {\rm In} and {\rm Ga} have smaller $ns-np$ and 
$ns-(n-1)d$ energy spacings as compared to group VA species ({\rm As} and {\rm P}).
The $s-p$  energy spacings are 5.7--6.4 eV  for
{\rm In} and {\rm Ga} and 9.3--8.4 eV for {\rm As} and {\rm P}, respectively. 

{\rm InP}, {\rm GaAs}, and {\rm InAs} are known to exist in 
several phases (See Refs.~\cite{RevModPhys.75.863, LB} and the references therein). 
The low pressure phase of all three 
({\rm GaAs, InP, InAs}) semiconductors is zinc-blende.  While there had been 
some discussions on the nature of the first high pressure phase of {\rm GaAs}, it is  now generally 
agreed that the first transition takes place at approximately 17 GPa and involves transition from 
zinc-blende ({\rm GaAs-I}) to orthorhombic $Cmcm$ phase ({\rm GaAs-II})~\cite{Nelmes1998}. 
The pressure release results in a transition
to a cinnabar phase followed by a transition to the original zinc-blende 
phase~\cite{McMahon1997}. No direct zinc-blende to cinnabar transition was observed. 
The first high-pressure phase in common ``cation" {\rm InP} and {\rm InAs} 
was experimentally found to be {\rm NaCl} phase~\cite{Vohra1985, Menoni1987, Nelmes1995}. 
The latter finding is supported by Density Functional calculations~\cite{Christensen1986, Zhang1987}. 

\begin{center}
\begin{figure}
\caption{\label{FIG:IIIVALC} LSDA error (\%) between KS DFT calculated lattice constant and experimental 
lattice constants at different ``levels of relativity" (FREL - fully relativistic, SREL - scalar relativistic, 
and NREL - nonrelativistic). LSDA lattice constants (FREL, SREL) 
underestimate experimental constants. In the case of {\rm InAs} and {\rm InP}, 
the NREL DFT lattice constant is larger than the experimental one which contradicts to the established 
trend. The experimental lattice constants
were determined at the room temperature.}
{\epsfig{file=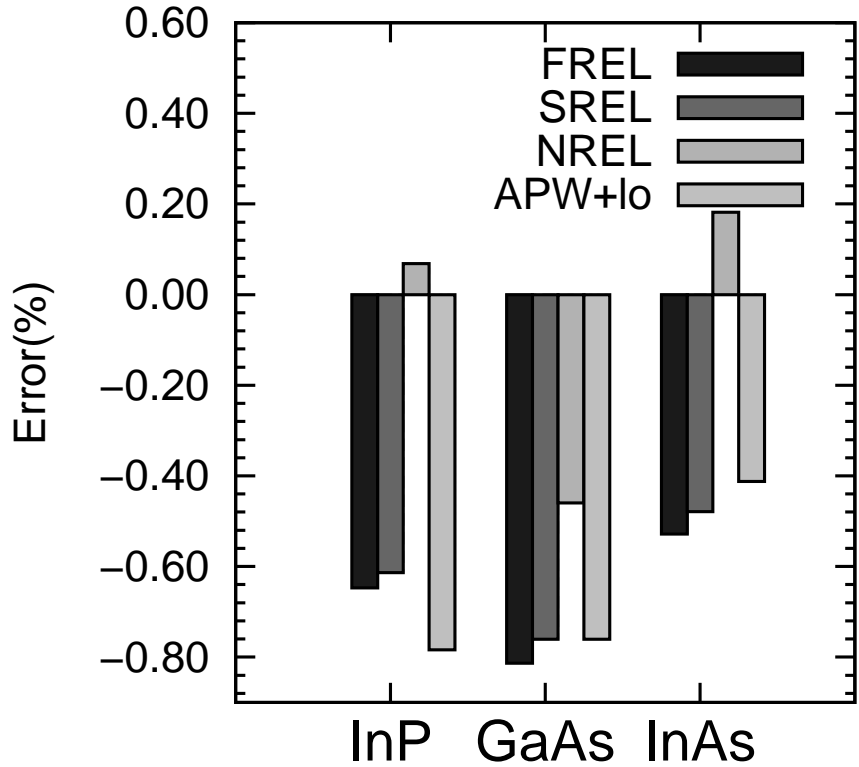,
width=0.7\textwidth,keepaspectratio,angle=270}}
\end{figure}
\end{center}

In connection with {\rm InAs} quantum dots in {\rm GaAs} or {\rm InP} matrix, 
the low pressure zinc-blende phase (Strukturbericht designation $B3$) is of primary interest. 
The structural parameters of {\rm InP}, {\rm GaAs}, and {\rm InAs} obtained from my Density Functional   
calculations with ZORA Relativistic Hamiltonian and APW+lo method are summarized in  Table~\ref{TAB:IIIVS}. 
Table~\ref{TAB:IIIVS} also shows experimental values~\cite{LB} which were measured
at room temperature and the results of APW+lo calculations. Figure~\ref{FIG:IIIVALC} shows the 
deviations between LSDA lattice constants and the experimental lattice constants. 

The finite temperature effects will increase the lattice constant.
Once these temperature effects are taken into account, the ``experimental" lattice constant is 
effectively reduced which will influence the conclusions about the accuracy of a given 
exchange-correlation functional. For example, in the case of {\rm GaAs}, the temperature effects 
lead to the increase in the lattice constant by 0.3\% from 5.638 \AA\ to  5.653 \AA\ \cite{Perdew2009}.
Since LSDA underestimates~\cite{vonBarth2004,Perdew2009} the bond lengths, 
the inclusion of finite temperature effects into 
consideration will improve the agreement between the theory and experiment. 

Table~\ref{TAB:IIIVS} and Figure~\ref{FIG:IIIVALC} show that
LSDA ZORA relativistic and scalar relativistic results underestimate 
the lattice constants by, approximately, 0.6\% ({\rm InP}), 0.8 \% ({\rm GaAs}),  and 
0.5 \% ({\rm InAs}).  The consideration 
of finite temperature effects will further improve agreement between the theory and 
experiment and it is likely that the error of LSDA relativistic ZORA calculation for lattice constants 
is within 0.5\%. 

In agreement with the established trend~\cite{Ziegler1981}, the relativity 
contracts the bond length. The DFT lattice constant decreases as the treatment  
of relativity changes from ``nonrelativistic" (NREL) to fully relativistic (FREL). 
Variational treatment of spin-orbital coupling does not seem to affect 
the structural properties by much, the reduction in 
the equilibrium lattice constant upon going from SREL to FREL is very small (within 0.05\%). 
It is important to include some kind of description of relativity for {\rm In} -- 
NREL lattice constants for {\rm InP} and {\rm InAs} are larger than experimental ones
which contradicts to the established LSDA trend~\cite{vonBarth2004,Perdew2009}.  

\begin{center}
\begin{figure}
\caption{\label{FIG:IIIVAB} LSDA bulk modulus calculated at different ``levels of relativity" 
(FREL - fully relativistic, SREL - scalar relativistic, 
and NREL - nonrelativistic). The LSDA calculations (this work) reproduce the 
``stiffness" trend $B_{InAs} < B_{InP} < B_{GaAs}$.}
{\epsfig{file=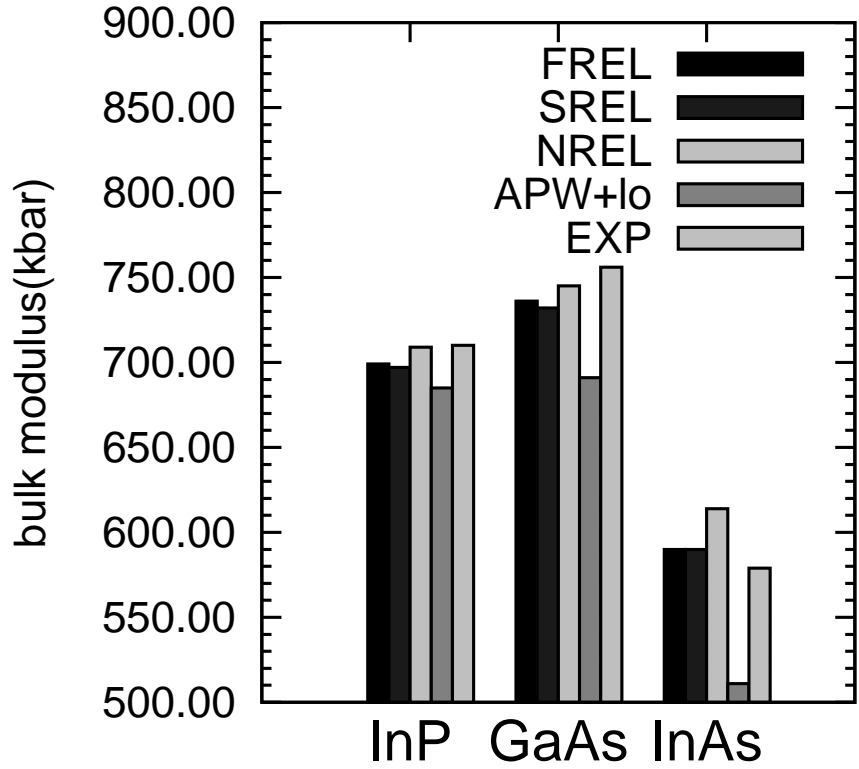,
width=0.7\textwidth,keepaspectratio,angle=270}}
\end{figure}
\end{center}

Figure~\ref{FIG:IIIVAB} shows LSDA bulk modulus calculated at different ``levels of relativity" 
(FREL - fully relativistic with variational treatment of spin orbital coupling, SREL - scalar relativistic, 
and NREL - nonrelativistic). My LSDA calculations reproduce the 
``stiffness" trend $B_{InAs} < B_{InP} < B_{GaAs}$. The bulk modulus decreases
upon going from NREL to SREL description and, then, slightly increases by going from SREL to FREL. 
In the case of {\rm InP} and {\rm GaAs}, the LSDA bulk modulus $B_{LSDA}$ is smaller (less stiffer) 
than the experimental bulk modulus. This seems to contradict to the ``established" LSDA trend of $B_{LSDA}$ being 
too stiff, note, however, that by examining Table V of Ref.~\cite{Perdew2009} one can not conclude this 
with respect to group IVA and group IIIA-VA semiconductors. In the case of {\rm InAs}, ZORA FREL and SREL 
bulk modulus is by 2\% more stiffer than the experimental bulk modulus. 

I find that my APW+lo calculations are in good agreement with the ZORA calculations. 
The  APW+lo lattice constants are within 0.2\% of the ZORA lattice constants. 
The bulk moduli obtained from the {\rm APW+lo} calculations 
are smaller than those obtained with BAND program (FREL) by 2 \%, 6\%, and 15\% 
for {\rm InP}, {\rm GaAs}, and {\rm InAs}, respectively, but the ``stiffness" trend $B_{InAs} < B_{InP} < B_{GaAs}$ 
is reproduced. 

\subsection{\label{S1} {\rm InP}, {\rm GaAs}, and {\rm InAs}: PDOS and OPWDOS Analysis} 

\begin{center}
\begin{figure}
\caption{\label{FIG:INPRD} PDOS in {\rm InP}. Results of fully relativistic calculations
for several lattice constants.}
{\epsfig{file=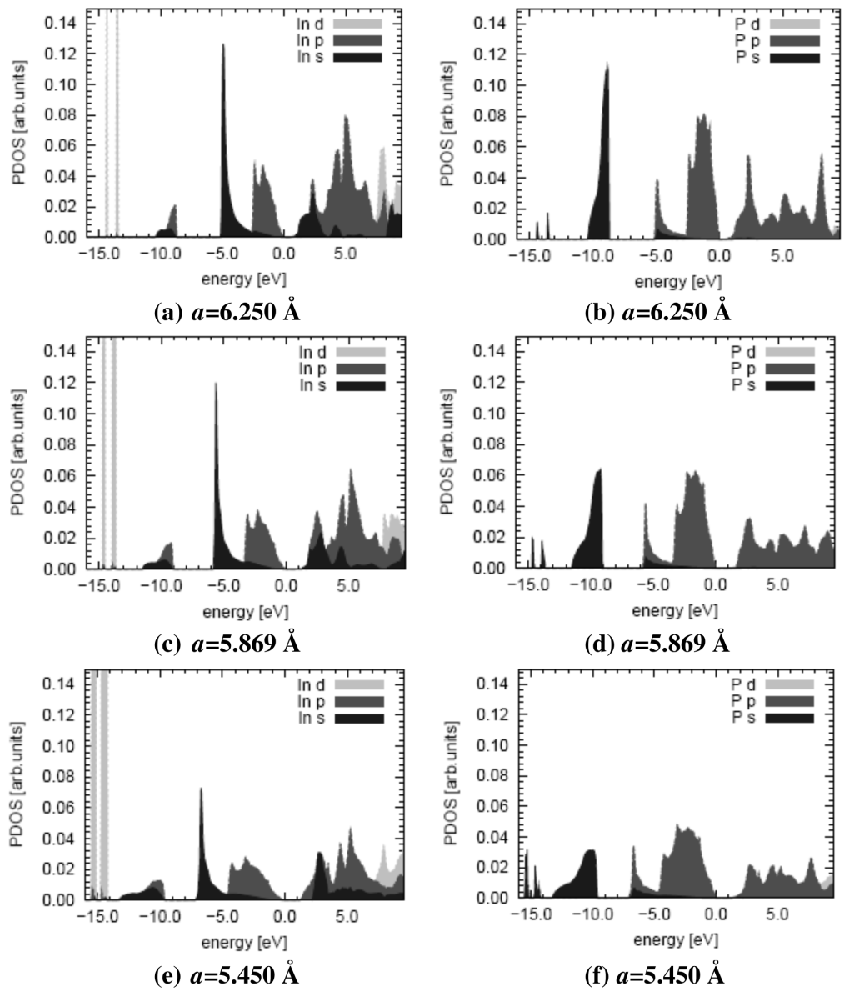,
width=1.0\textwidth,keepaspectratio,angle=0}}
\end{figure}
\end{center}

\begin{center}
\begin{figure}
\caption{\label{FIG:GAASRD} PDOS in {\rm GaAs}. Results of fully relativistic calculations
for several lattice constants. }
{\epsfig{file=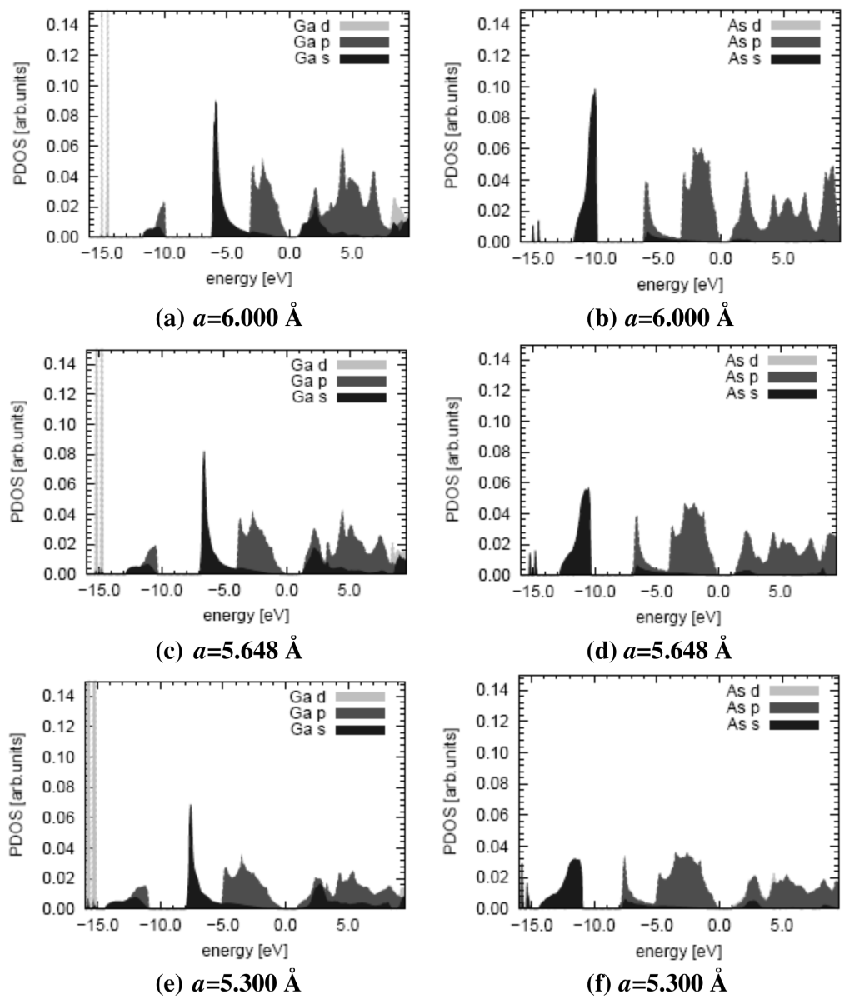,
width=1.0\textwidth,keepaspectratio,angle=0}}
\end{figure}
\end{center}

\begin{center}
\begin{figure}
\caption{\label{FIG:INASRD} PDOS in {\rm InAs}. Results of fully relativistic calculations
for several lattice constants.}
{\epsfig{file=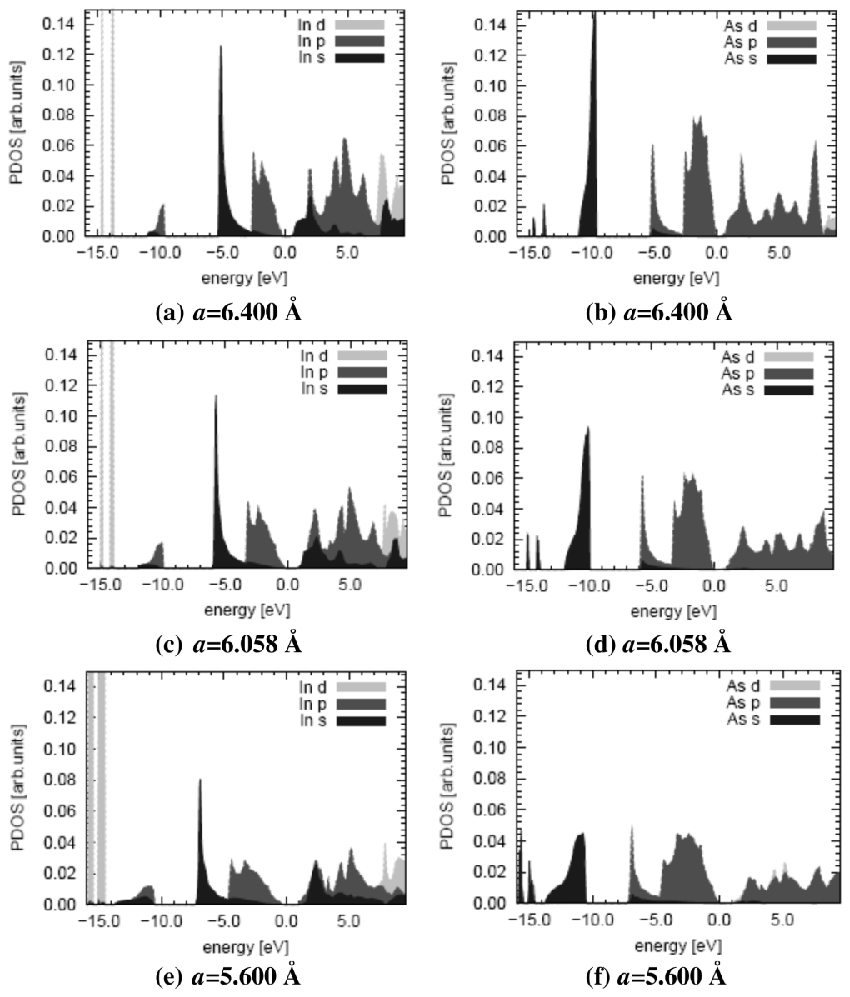,
width=1.0\textwidth,keepaspectratio,angle=0}}
\end{figure}
\end{center}

Figs.~\ref{FIG:INPRD},~\ref{FIG:GAASRD}, and~\ref{FIG:INASRD}
show relativistic PDOS for {\rm InP}, {\rm  GaAs}, and {\rm InAs}, 
respectively. The PDOS is calculated at three values of the lattice constant corresponding to the
tensile hydrostatic strain (Figs.~(a) and~(b)), equilibrium (Figs.~(c) and~(d)), and 
compressive hydrostatic strain (Figs.~(e) and~(f)).

Let us consider PDOS at the equilibrium (experimental) lattice constants 
in {\rm InP}, {\rm GaAs}, and {\rm InAs}~(middle rows of 
Figs.~\ref{FIG:INPRD},~\ref{FIG:GAASRD}, and~\ref{FIG:INASRD}). 
The left~(c) and right~(d) columns show PDOS on the cation ({\rm In}, {\rm Ga}) 
and anion ({\rm As}, {\rm P}) atomic orbitals (AOs), respectively. The names ``cation" 
and ``anion" reflect the move away from covalent bonding and towards ionicity 
in {\rm InP, GaAs}, and {\rm InAs} semiconductors. The Hirshfeld charge analysis~\cite{Hirshfeld} performed 
in this work reveals that electron density transfers from regions near {\rm In} and 
{\rm Ga} ions into regions near {\rm As} and {\rm P} ions making
{\rm In} and {\rm Ga}  ``positively"  and 
{\rm As} and {\rm P}  ``negatively" charged, respectively. 

We find that PDOS of {\rm InP}, {\rm GaAs}, and  {\rm InAs} in the valence band 
energy region (up to 16 eV below the Fermi energy, the Fermi energy is at zero) 
consists of four main ``spectral features".

The first ``spectral feature" in the PDOS is a  peak of  
broad character just below the Fermi energy. The width of this peak is, approximately, 
3.2 eV for {\rm InAs} and {\rm InP} and  4.4 eV for {\rm GaAs}. The main 
contribution to this ``spectral feature" stems from anion and, to a lesser extent, from cation 
atomic orbitals (AOs) of $p$ type. There is a small contribution from  $s$ and $d$ cation orbitals 
and, in the case of {\rm GaAs} and {\rm InAs}, $s$ and $d$ anion orbitals. 

The second feature is a very sharp peak centered around 5.4 eV ({\rm InAs},~{\rm InP}) 
and 6.5 eV (\rm GaAs) below the Fermi energy. The main contributions to this peak are 
cation AOs of $s$ type. The smaller contributions to this peak are arising from anion and cation AOs of $p$ 
type. The second feature
has a slowly decaying tail which contributes to the very top of the valence band.

The third peak in the PDOS is located, approximately, at 9.5 eV ({\rm InP}), 12.2  eV ({\rm GaAs}),
and 10.9 eV (\rm {InAs}) below the Fermi energy. The main contribution to this peak stem from anion
$s$ type orbitals as well as  from cation $p$, $s$ and $d$  AOs.
The last ``spectral feature" is a very sharp peak in the PDOS due to the cation $d$ orbitals and 
anion $s$ type orbitals. 

The PDOS in the conduction band energy region is shown up to 9.5 eV above the Fermi energy and 
consists of several closely spaced peaks. The main contributions in this energy window are the anion
$p$ and $d$ and cation $p$, $s$, and $d$ AOs. The PDOS does not show significant anion $s$-type 
contributions in the conduction band region. 
 
Figs.~(a),(b) and (c),(d) demonstrate the effect of the hydrostatic tensile and compressive strains on the PDOS. 
The compressive hydrostatic strain 1) broadens PDOS peaks; 2) decreases 
magnitude of the PDOS peaks; 3) changes the relative position of the peaks. The 
``general structure" of the PDOS is preserved.

\begin{center}
\begin{figure}
\caption{\label{FIG:INPSRO} OPWDOS in {\rm InP} ($a=5.869$~\AA). Results of scalar relativistic calculations.}
{\epsfig{file=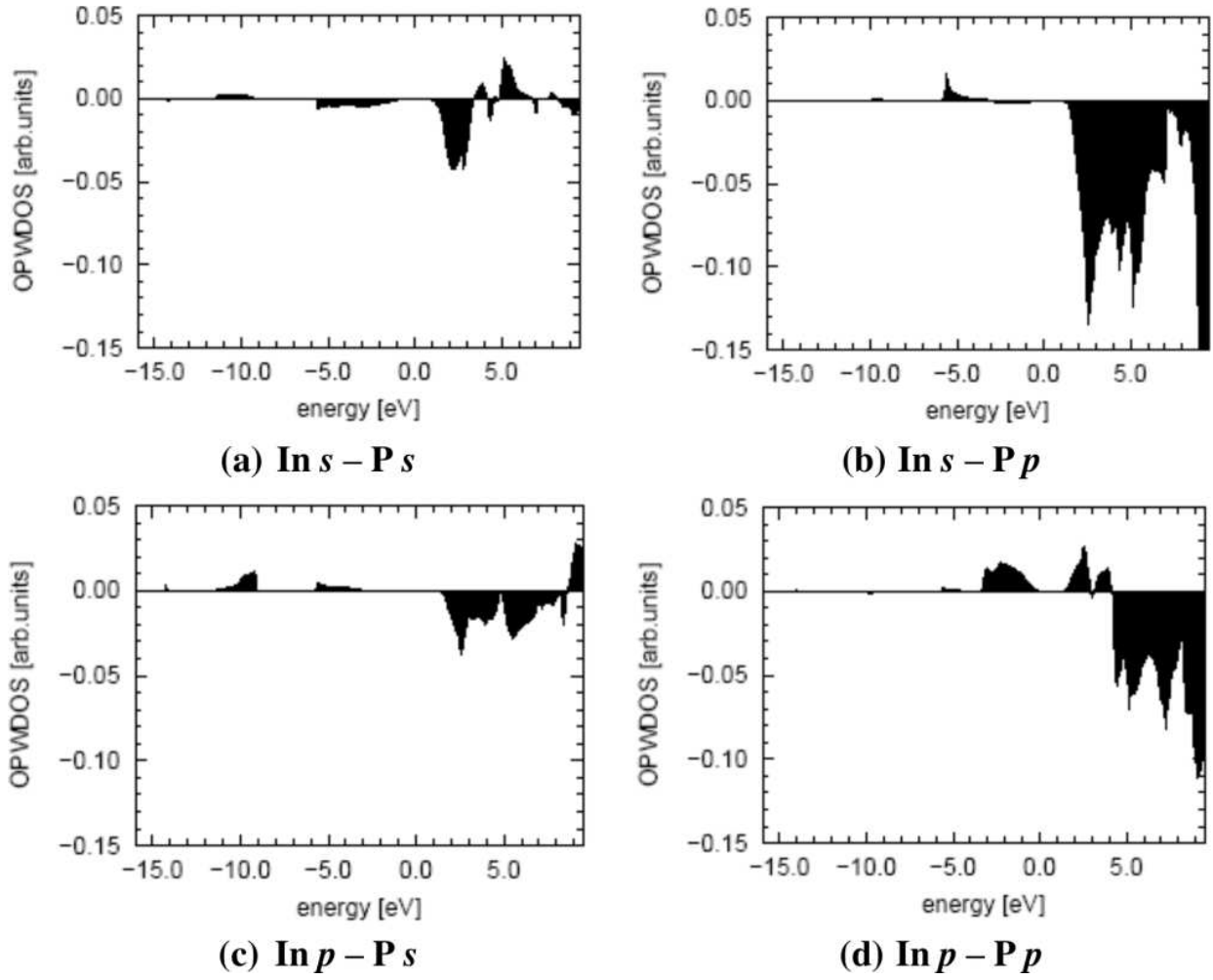,
width=1.0\textwidth,keepaspectratio,angle=0}}
\end{figure}
\end{center}

\begin{center}
\begin{figure}
\caption{\label{FIG:GAASSRO} OPWDOS in {\rm GaAs} ($a=5.648$~\AA). Results of scalar relativistic calculations.}
{\epsfig{file=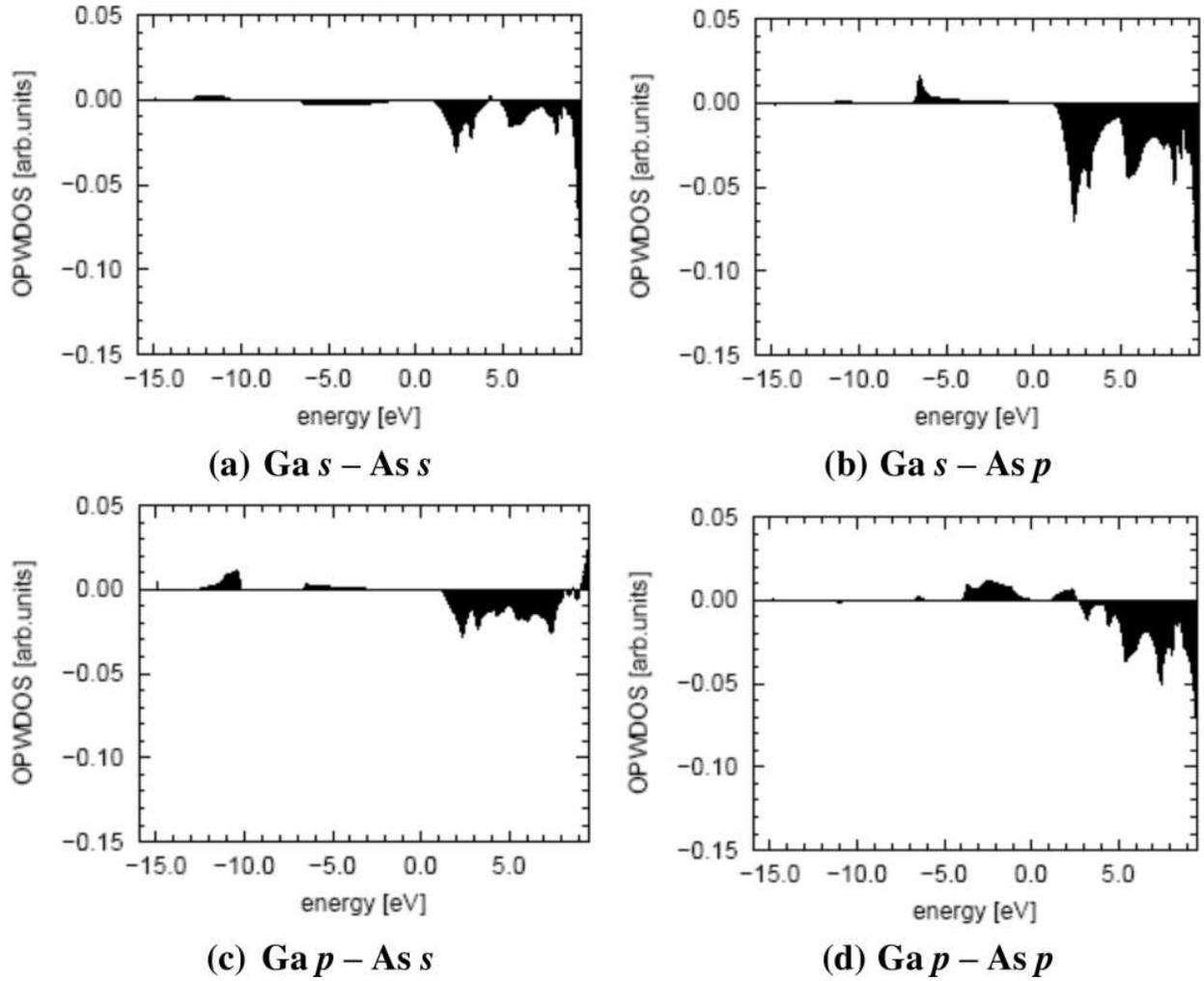,
width=1.0\textwidth,keepaspectratio,angle=0}}
\end{figure}
\end{center}

\begin{center}
\begin{figure}
\caption{\label{FIG:INASSRO} OPWDOS in {\rm InAs} ($a=6.058$~\AA). Results of scalar relativistic calculations.}
{\epsfig{file=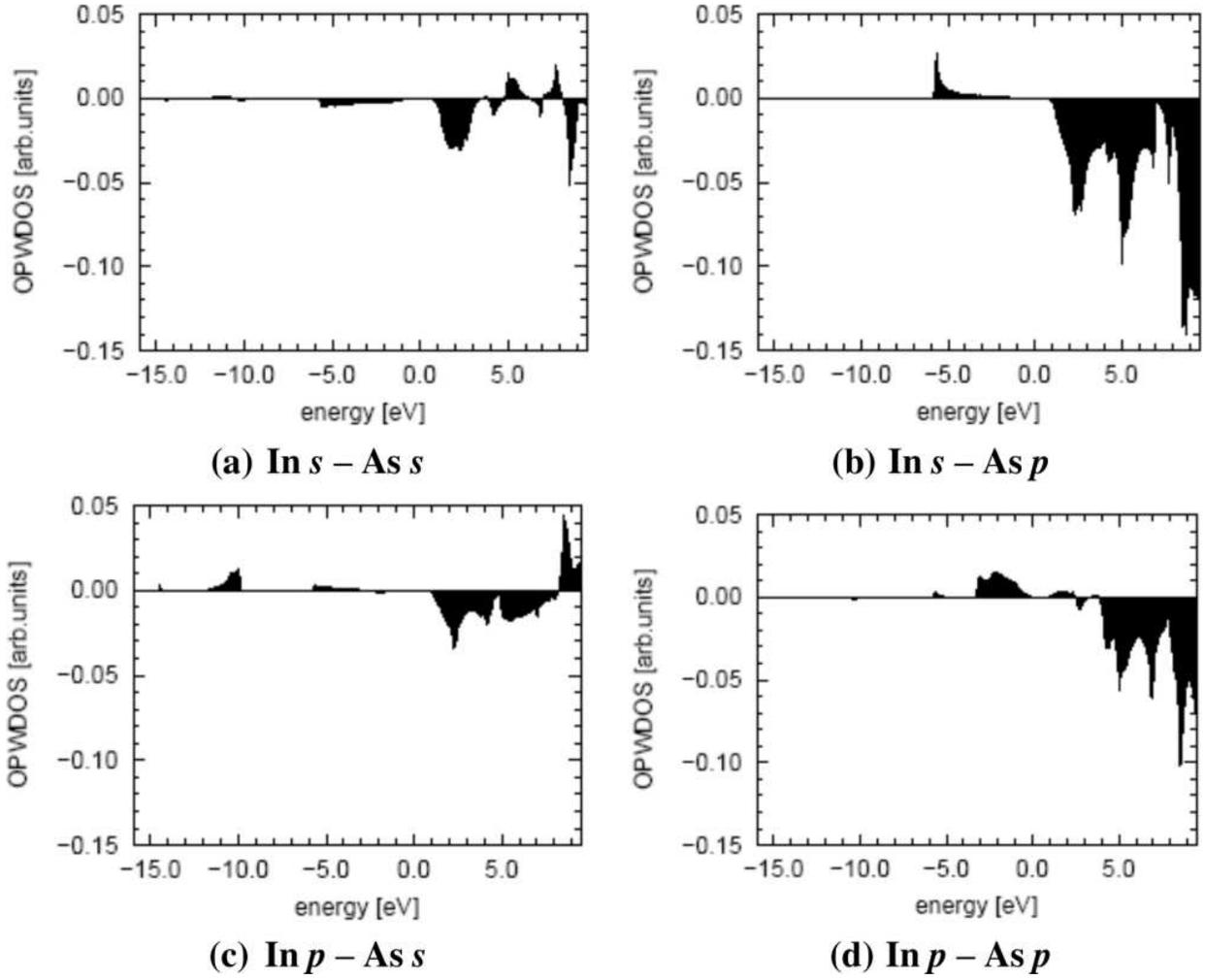,
width=1.0\textwidth,keepaspectratio,angle=0}}
\end{figure}
\end{center}

Figs.~\ref{FIG:INPSRO},~\ref{FIG:GAASSRO}, and~\ref{FIG:INASSRO} 
show OPWDOS for {\rm InP}, {\rm GaAs},  and {\rm InAs}, respectively. The OPWDOS is
calculated at the experimental equilibrium lattice constants and 
provides a clear pictorial representation of orbital interactions 
in IIIA-VA {\rm InP}, {\rm GaAs},  and {\rm InAs} zinc-blende semiconductors.

The OPWDOS in the valence band energy region is of mostly bonding character. 
The OPWDOS  shows that the very top of the valence band 
is strongly stabilized by cation-anion
$p-p$ orbital interactions. The energy region corresponding to the second PDOS peak 
is characterized by the cation $s$ - anion $p$ and cation $p$ - anion $s$ bonding interactions 
as well as a slightly anti-bonding cation $s$ - anion $s$ interactions. The energy region corresponding to 
the third PDOS peak is characterized by the cation $p$ - anion $s$ interaction.

The OPWDOS in the conduction band energy region is mostly of the anti-bonding character 
and has peaks that are generally higher in magnitude than the OPWDOS peaks in the valence band 
region. The latter is a demonstration of a well known fact that 
the ``anti-bonding is more anti-bonding than bonding is bonding" (see Ref.~\cite{Kadantsev2008} and the references 
therein).  The bottom of the conduction 
band is strongly ``destabilized" by cation-anion $s-p$, $p-s$, and $s-s$ 
interactions. It is interesting to note that the cation-anion $p-p$ interaction for the very bottom of 
the conduction band (within 1 eV) is bonding. At the higher energies,  the $p-p$ interaction
becomes strongly anti-bonding. 

\subsection{{\rm InP}, {\rm GaAs}, and {\rm InAs}: Electronic Structure Parameters} 

The PDOS and OPWDOS presented in Section~\ref{S1} describe the electronic structure qualitatively. 
Tables~\ref{TAB:IIIVE} and~\ref{TAB:IIIVDEFPOT} show  results for the parameters of the electronic structure
for {\rm InP}, {\rm GaAs}, and {\rm InAs} IIIA-VA zinc-blende semiconductors. 
These parameters are the band gaps for the transitions between the
high-symmetry points of the Brillouin zone $E_{gap}^{1 \rightarrow 2} = E_2 - E_1$, split-off energies 
at the $\Gamma$ and $L$ points of the Brillouin zone, position (with respect to the Fermi energy) 
and width of the cation $d$-band at the $\Gamma$ point, and the width of the upper part of the valence band. 
Table~\ref{TAB:IIIVDEFPOT} summarizes the 
relative volume deformation potentials for a specific $1 \rightarrow 2$ transition
defined as
\begin{equation}
a_V^{1 \rightarrow 2} = \frac{dE_{21}}{d ln V}.
\end{equation}	
The electronic band structure parameters are computed at several levels 
of relativity (FREL - fully relativistic, SREL - scalar relativistic, 
and NREL - nonrelativistic) as well as with the APW+lo method (fully relativistic approach). The 
results of these calculations are compared with available experimental data. 

In agreement with previous work, Table~\ref{TAB:IIIVE} shows that LSDA KS DFT severely 
underestimates band gaps. The small LSDA band gap becomes even smaller when one introduces 
explicit description of relativity. For example, in the case of {\rm GaAs}, the direct gap
$E_{gap}^{\Gamma_V \rightarrow \Gamma_C}$ decreases from 1.0 eV to 0.4 eV as the 
``level of relativity" changes from nonrelativistic to fully relativistic. The relativity 
especially strongly affects band gaps at $\Gamma$ and $L$ points of the Brillouin zone.

In the Table~\ref{TAB:DFTRES}, I summarized orbital populations for three top valence bands 
(split-off $\Gamma_{7v}$, light-hole, and heavy hole $\Gamma_{8v}$) and the lowest conduction 
band ($\Gamma_{6c}$) at the high-symmetry points of the Brillouin zone. 
From Table~\ref{TAB:DFTRES}, one can see that the bottom of the conduction band at the $\Gamma$ 
and $L$ points has significant contributions from cation $s$-type AOs, whereas at the $X$ point 
the bottom of the conduction band is made up of cation $p$-type orbitals. The reason 
for the strong ``relativistic" band gap decrease at the $\Gamma$ and $L$ points is that 
the conduction $s$ states are stabilized by relativity stronger than the valence $p$ states. 
The stabilization itself stems from the relativistic ``contraction" of the 
atomic orbitals~\cite{Pyykko1988}. Overall, the gaps in the relativistic description 
decrease substantially which might affect the conclusions with respect to the accuracy of a given 
exchange-correlational functional for the band gap calculation. 

For a fixed lattice constant, LSDA also poorly describes the energy differences within the conduction band. 
For example, $E(X_C)-E(L_C)$ energy difference is 508 meV from ZORA FREL calculations, whereas the experimental value
is 160 meV. The relative position of the conduction band minima is described only qualitatively 
$E_{gap}^{\Gamma_V \rightarrow \Gamma_C} < E_{gap}^{\Gamma_V \rightarrow L_C} < E_{gap}^{\Gamma_V \rightarrow X_C}$.

The strong underestimation of the band gaps may result in the wrong energetic order 
of bands in some specific points of the Brillouin zone. This happens at the $\Gamma$ point 
for {\rm InAs} at the equilibrium lattice constant, 
where a conduction band $\Gamma_{6c}$ is strongly stabilized and lies below 
the split-off band $\Gamma_{7v}$.  The strain also may affect the energetic order of bands. 

\begin{center}
\begin{figure}
\caption{\label{FIG:SRELBS} Band Structure of {\rm InP}, {\rm GaAs}, and {\rm InAs}. 
Results of scalar relativistic calculations. }
{\epsfig{file=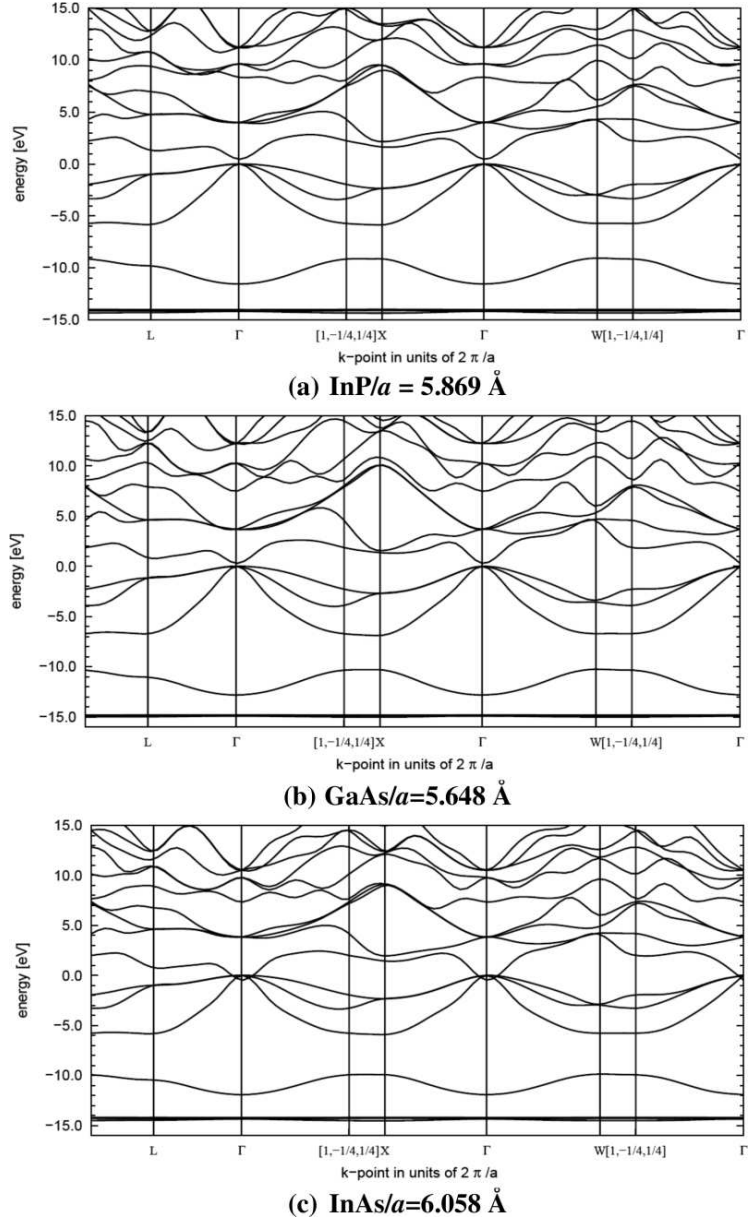,
width=0.6\textwidth,keepaspectratio,angle=0}}
\end{figure}
\end{center}

\begin{center}
\begin{figure}
\caption{\label{FIG:FRELBS} Band Structure of {\rm InP}, {\rm GaAs}, and {\rm InAs}. 
Results of fully relativistic calculations.}
{\epsfig{file=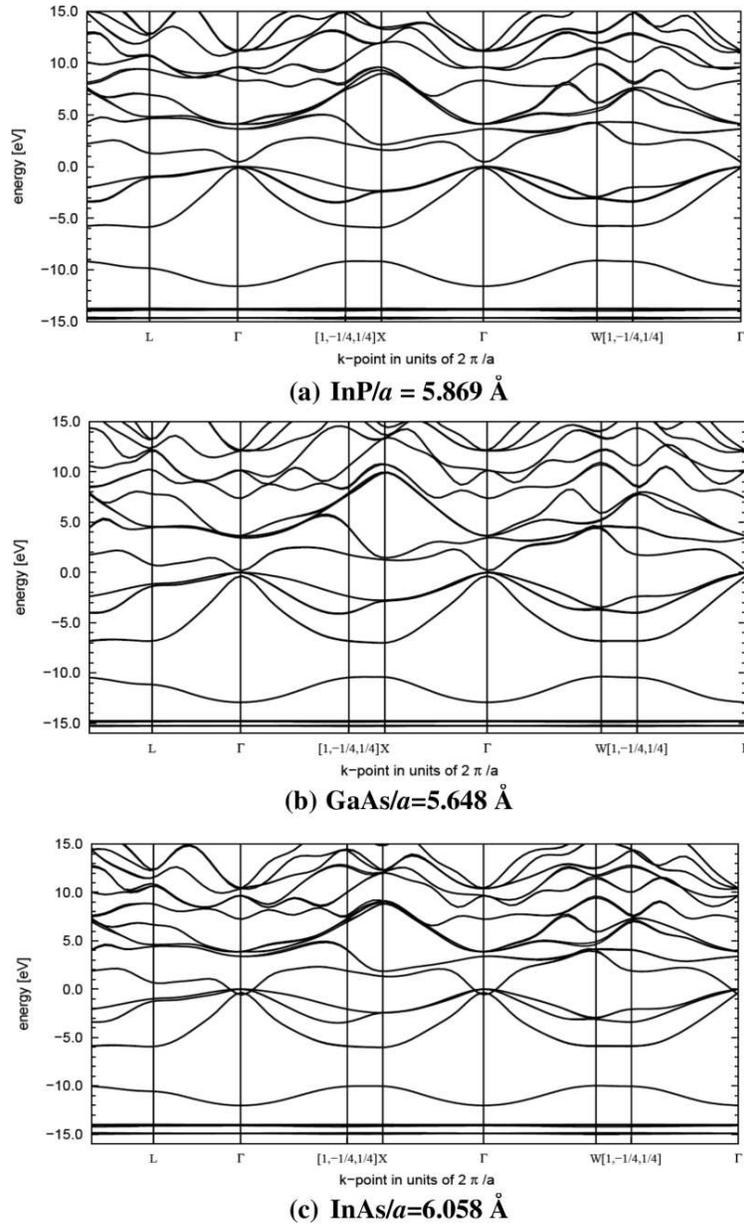,
width=0.6\textwidth,keepaspectratio,angle=0}}
\end{figure}
\end{center}

Figures~\ref{FIG:SRELBS} and~\ref{FIG:FRELBS} show scalar relativistic and fully relativistic 
band structure plots for {\rm InP}, {\rm GaAs}, and {\rm InAs}, respectively. The band structure 
is computed along the edges connecting the high-symmetry points of the Brillouin zone. The Cartesian 
coordinates of these 
high-symmetry points are  summarized in Table~\ref{TAB:KPOINTS}. From these band structure 
plots one can see the band crossing  at the $\Gamma$-point  for {\rm InAs} at the equilibrium
lattice constant. This band crossing also occurs in {\rm InP} and {\rm GaAs} subjected 
to the tensile hydrostatic strain. 

It is found that the split-off energies at the ${\Gamma}$ and $L$ points 
are reproduced very accurately within ZORA fully relativistic approach, the agreement with 
the experiment is a few \% or several meVs. The fully relativistic treatment for the calculation 
of the split-off energies is essential as both scalar relativistic/nonrelativistic calculations 
lead to the six-fold (including spin) degeneracy of the valence band at the $\Gamma$-point (Fig.~\ref{FIG:SRELBS}).

The relative volume deformation potential describes how fast a given band gap changes 
with volume. The negative (positive) relative deformation 
potential (in our definition) means that the band gap increases (decreases) as volume decreases. 
Table~\ref{TAB:IIIVE} shows 
that for both $\Gamma_V \rightarrow \Gamma_C$ and $\Gamma_V \rightarrow L_C$  transitions, 
both gaps are increasing as volume decreases (negative deformation potential), whereas for the
$\Gamma_V \rightarrow X_C$ transition, the band 
gap decreases (the deformation potential is positive). The differences in the sign of the deformation 
potential for $\Gamma_V \rightarrow \Gamma_C$ and $\Gamma_V \rightarrow L_C$ transitions on one hand and 
$\Gamma_V \rightarrow X_C$ transition on the other hand are attributed to the ``different nature" of the conduction 
band minimum at these points (see Table~\ref{TAB:DFTRES}). 

The calculated absolute magnitude of 
the ``rate of change" in the gap is the largest for {\rm GaAs} and decreases for semiconductors with a larger 
lattice constant ({\rm InP}, {\rm InAs}). The experimental relative deformation potentials are obtained from the 
direct band gap pressure dependence coefficients and experimental bulk moduli. The 
experimental trend in the absolute magnitude of the ``rate of change" in the gap 
is {\rm GaAs}, {\rm InAs}, and {\rm InP}. Note, however, that experimental uncertainties 
for the relative deformation potential can be as large as 1 eV. 
The relative deformation potentials are quite close for fully relativistic and scalar 
relativistic calculations and, therefore, a fully relativistic calculation of this quantity 
does not seem essential, at least, for the transitions considered in this work. 

Finally, there are  two other quantities which sensitively depend on fully relativistic calculation -- 
the cation $d$-band width and the width of the upper part of the valence band UVBW. Both band widths
increase as the treatment of relativity changes from non-relativistic to fully relativistic level. For  
UVBW, the increase is 6\%-8\% (0.4-0.5 eV). The $d$-band width increases dramatically 
from 0.10-0.15 eV to 0.9 eV.  

The agreement between ZORA fully relativistic and APW+lo calculations is 
exceptionally good, especially, for the relative deformation potentials, $E_d^{\Gamma}$, 
$\delta E_d^{\Gamma}$, and for UVBW. The split-off energies are reproduced 
within 10 meVs, and the gaps usually agree within 20 meVs. 

\section{Conclusions}

First-principles full potential calculations based on the Zeroth Order Regular 
Approximation (ZORA) relativistic Hamiltonian and the Kohn-Sham form of Density 
Functional Theory (KS DFT) were reported for group 
IIIA-VA ({\rm InAs}, {\rm GaAs}, {\rm InP})  semiconductors. 
The effects of relativity were elucidated by performing fully 
relativistic, scalar relativistic, and nonrelativistic 
calculations. The inclusion of relativity  led 
to the contraction of the bond length, strong stabilization of the 
conduction band at ${\Gamma}$ and $L$ points of the Brillouin zone, 
and broadening of the upper part of the valence band. The inclusion of relativity at least on 
the scalar relativistic level was found to be essential for the 
accurate calculation of structural properties. Electronic band structure parameters 
were determined including split-off energies, band gaps,  deformation potentials, and 
populations at the high-symmetry points of the Brillouin zone. 
It was found that the split-off energies can be determined with very good accuracy. 
In agreement with previous work, LSDA KS DFT severely underestimates band gaps which 
may result in the wrong energetic order of bands at specific points of the reciprocal space. 
It was found that relativistic LSDA describes the sequence of conduction band minima 
only qualitatively. The relative band gap deformation potentials were determined and
compared with the available experimental data. The relativistic LSDA relative deformation 
potentials at the $\Gamma$ point were found to be too small. 
The nature of the chemical bonding at the equilibrium and under hydrostatic 
strain was investigated using projected  (PDOS) and overlap population 
weighted density of states (OPWDOS). It was found that OPWDOS in the 
valence and conduction band energy regions is of mostly bonding and anti-bonding 
type, respectively.  ZORA results were compared with Augmented Plane 
Wave plus Local Orbitals method (APW+lo) and a good agreement between 
the two sets of calculations was established . Viability and 
robustness of the ZORA Hamiltonian for the investigation of electronic and 
structural properties of semiconductors was reaffirmed. 

\section{Acknowledgment}
The author acknowledges support by the NRC-NSERC-BDC Nanotechnology  project, 
QuantumWorks, NRC-CNRS CRP and CIFAR. The author thanks Prof. Pawel Hawrylak 
for stimulating discussions. The author thanks Dr. Maxim Shishkin for 
reading the manuscript and making useful suggestions. 


\begin{thebibliography}{58}
\expandafter\ifx\csname natexlab\endcsname\relax\def\natexlab#1{#1}\fi
\expandafter\ifx\csname bibnamefont\endcsname\relax
  \def\bibnamefont#1{#1}\fi
\expandafter\ifx\csname bibfnamefont\endcsname\relax
  \def\bibfnamefont#1{#1}\fi
\expandafter\ifx\csname citenamefont\endcsname\relax
  \def\citenamefont#1{#1}\fi
\expandafter\ifx\csname url\endcsname\relax
  \def\url#1{\texttt{#1}}\fi
\expandafter\ifx\csname urlprefix\endcsname\relax\def\urlprefix{URL }\fi
\providecommand{\bibinfo}[2]{#2}
\providecommand{\eprint}[2][]{\url{#2}}

\bibitem[{\citenamefont{Jacak et~al.}(1998)\citenamefont{Jacak, Hawrylak, and
  Wojs}}]{Hawrylak1997}
\bibinfo{author}{\bibfnamefont{L.}~\bibnamefont{Jacak}},
  \bibinfo{author}{\bibfnamefont{P.}~\bibnamefont{Hawrylak}}, \bibnamefont{and}
  \bibinfo{author}{\bibfnamefont{A.}~\bibnamefont{Wojs}},
  \emph{\bibinfo{title}{Quantum Dots}} (\bibinfo{publisher}{Springer-Verlag},
  \bibinfo{address}{Berlin}, \bibinfo{year}{1998}).

\bibitem[{\citenamefont{Dalacu et~al.}(2009)\citenamefont{Dalacu, Frederick,
  Kim, Reimer, Lapointe, Poole, Aers, Williams, McKinnon, Korkusinski, 
  and Hawrylak}}]{Dalacu2009}
\bibinfo{author}{\bibfnamefont{D.}~\bibnamefont{Dalacu}},
  \bibinfo{author}{\bibfnamefont{M.~E.} \bibnamefont{Reimer}},
  \bibinfo{author}{\bibfnamefont{S.}~\bibnamefont{Frederick}},
  \bibinfo{author}{\bibfnamefont{D.}~\bibnamefont{Kim}},
  \bibinfo{author}{\bibfnamefont{J.}~\bibnamefont{Lapointe}},
  \bibinfo{author}{\bibfnamefont{P.~J.} \bibnamefont{Poole}},
  \bibinfo{author}{\bibfnamefont{G.~C.} \bibnamefont{Aers}},
  \bibinfo{author}{\bibfnamefont{R.~L.} \bibnamefont{Williams}},
  \bibinfo{author}{\bibfnamefont{W.~R.} \bibnamefont{McKinnon}},
  \bibinfo{author}{\bibfnamefont{M.}~\bibnamefont{Korkusinski}},
  \bibnamefont{and} \bibinfo{author}{\bibfnamefont{P.}~\bibnamefont{Hawrylak}}, 
  \bibinfo{journal}{Laser and Photonics Reviews}
  \textbf{\bibinfo{volume}{4}}, \bibinfo{pages}{283}
  (\bibinfo{year}{2010}).

\bibitem[{\citenamefont{Shields}(2007)}]{Shields2007}
\bibinfo{author}{\bibfnamefont{A.~J.} \bibnamefont{Shields}},
  \bibinfo{journal}{Nature Photonics} \textbf{\bibinfo{volume}{1}},
  \bibinfo{pages}{215} (\bibinfo{year}{2007}).

\bibitem[{\citenamefont{Stevenson et~al.}(2006)\citenamefont{Stevenson, Young,
  Atkinson, Cooper, Ritchie, and Shields}}]{Stevenson2006}
\bibinfo{author}{\bibfnamefont{R.~M.} \bibnamefont{Stevenson}},
  \bibinfo{author}{\bibfnamefont{R.~J.} \bibnamefont{Young}},
  \bibinfo{author}{\bibfnamefont{P.}~\bibnamefont{Atkinson}},
  \bibinfo{author}{\bibfnamefont{K.}~\bibnamefont{Cooper}},
  \bibinfo{author}{\bibfnamefont{D.~A.} \bibnamefont{Ritchie}},
  \bibnamefont{and} \bibinfo{author}{\bibfnamefont{A.~J.}
  \bibnamefont{Shields}}, \bibinfo{journal}{Nature}
  \textbf{\bibinfo{volume}{439}}, \bibinfo{pages}{179} (\bibinfo{year}{2006}).

\bibitem[{\citenamefont{Akopian et~al.}(2006)\citenamefont{Akopian, Lindner,
  Poem, Berlatzky, Avron, Gershoni, Gerardot, and Petroff}}]{Akopian2006}
\bibinfo{author}{\bibfnamefont{N.}~\bibnamefont{Akopian}},
  \bibinfo{author}{\bibfnamefont{N.~H.} \bibnamefont{Lindner}},
  \bibinfo{author}{\bibfnamefont{E.}~\bibnamefont{Poem}},
  \bibinfo{author}{\bibfnamefont{Y.}~\bibnamefont{Berlatzky}},
  \bibinfo{author}{\bibfnamefont{J.}~\bibnamefont{Avron}},
  \bibinfo{author}{\bibfnamefont{D.}~\bibnamefont{Gershoni}},
  \bibinfo{author}{\bibfnamefont{B.~D.} \bibnamefont{Gerardot}},
  \bibnamefont{and} \bibinfo{author}{\bibfnamefont{P.~M.}
  \bibnamefont{Petroff}}, \bibinfo{journal}{Phys. Rev. Lett.}
  \textbf{\bibinfo{volume}{96}}, \bibinfo{pages}{130501}
  (\bibinfo{year}{2006}).

\bibitem[{\citenamefont{Greilich et~al.}(2006)\citenamefont{Greilich, Schwab,
  Berstermann, Auer, Oulton, Yakovlev, Bayer, Stavarache, Reuter, and
  Wieck}}]{Greilich2006}
\bibinfo{author}{\bibfnamefont{A.}~\bibnamefont{Greilich}},
  \bibinfo{author}{\bibfnamefont{M.}~\bibnamefont{Schwab}},
  \bibinfo{author}{\bibfnamefont{T.}~\bibnamefont{Berstermann}},
  \bibinfo{author}{\bibfnamefont{T.}~\bibnamefont{Auer}},
  \bibinfo{author}{\bibfnamefont{R.}~\bibnamefont{Oulton}},
  \bibinfo{author}{\bibfnamefont{D.~R.} \bibnamefont{Yakovlev}},
  \bibinfo{author}{\bibfnamefont{M.}~\bibnamefont{Bayer}},
  \bibinfo{author}{\bibfnamefont{V.}~\bibnamefont{Stavarache}},
  \bibinfo{author}{\bibfnamefont{D.}~\bibnamefont{Reuter}}, \bibnamefont{and}
  \bibinfo{author}{\bibfnamefont{A.}~\bibnamefont{Wieck}},
  \bibinfo{journal}{Phys. Rev. B} \textbf{\bibinfo{volume}{73}},
  \bibinfo{pages}{045323} (\bibinfo{year}{2006}).

\bibitem[{\citenamefont{Hohenberg and Kohn}(1964)}]{Hohenberg1964}
\bibinfo{author}{\bibfnamefont{P.}~\bibnamefont{Hohenberg}} \bibnamefont{and}
  \bibinfo{author}{\bibfnamefont{W.}~\bibnamefont{Kohn}},
  \bibinfo{journal}{Phys. Rev.} \textbf{\bibinfo{volume}{136}},
  \bibinfo{pages}{B864} (\bibinfo{year}{1964}).

\bibitem[{\citenamefont{Kohn and Sham}(1965)}]{Kohn1965}
\bibinfo{author}{\bibfnamefont{W.}~\bibnamefont{Kohn}} \bibnamefont{and}
  \bibinfo{author}{\bibfnamefont{L.~J.} \bibnamefont{Sham}},
  \bibinfo{journal}{Phys. Rev.} \textbf{\bibinfo{volume}{140}},
  \bibinfo{pages}{A1133} (\bibinfo{year}{1965}).

\bibitem[{\citenamefont{von Barth and Hedin}(1972)}]{vonBarth1972}
\bibinfo{author}{\bibfnamefont{U.}~\bibnamefont{von Barth}} \bibnamefont{and}
  \bibinfo{author}{\bibfnamefont{L.}~\bibnamefont{Hedin}}, \bibinfo{journal}{J.
  Phys. C: Solid State Phys.} \textbf{\bibinfo{volume}{5}}
  (\bibinfo{year}{1972}).

\bibitem[{\citenamefont{Mujica et~al.}(2003)\citenamefont{Mujica, Rubio,
  Mu\~noz, and Needs}}]{RevModPhys.75.863}
\bibinfo{author}{\bibfnamefont{A.}~\bibnamefont{Mujica}},
  \bibinfo{author}{\bibfnamefont{A.}~\bibnamefont{Rubio}},
  \bibinfo{author}{\bibfnamefont{A.}~\bibnamefont{Mu\~noz}}, \bibnamefont{and}
  \bibinfo{author}{\bibfnamefont{R.~J.} \bibnamefont{Needs}},
  \bibinfo{journal}{Rev. Mod. Phys.} \textbf{\bibinfo{volume}{75}},
  \bibinfo{pages}{863} (\bibinfo{year}{2003}).

\bibitem[{\citenamefont{Van~de Walle}(1989)}]{VandeWalle1989}
\bibinfo{author}{\bibfnamefont{C.~G.} \bibnamefont{Van~de Walle}},
  \bibinfo{journal}{Phys. Rev. B} \textbf{\bibinfo{volume}{39}},
  \bibinfo{pages}{1871} (\bibinfo{year}{1989}).

\bibitem[{\citenamefont{Wei and Zunger}(1998)}]{wei1998}
\bibinfo{author}{\bibfnamefont{S.-H.} \bibnamefont{Wei}} \bibnamefont{and}
  \bibinfo{author}{\bibfnamefont{A.}~\bibnamefont{Zunger}},
  \bibinfo{journal}{Appl. Phys. Lett.} \textbf{\bibinfo{volume}{72}},
  \bibinfo{pages}{2011} (\bibinfo{year}{1998}).

\bibitem[{\citenamefont{Wei and Zunger}(1999)}]{wei1999}
\bibinfo{author}{\bibfnamefont{S.-H.} \bibnamefont{Wei}} \bibnamefont{and}
  \bibinfo{author}{\bibfnamefont{A.}~\bibnamefont{Zunger}},
  \bibinfo{journal}{Phys. Rev. B} \textbf{\bibinfo{volume}{60}},
  \bibinfo{pages}{5404} (\bibinfo{year}{1999}).

\bibitem[{\citenamefont{Li et~al.}(2006)\citenamefont{Li, Gong, and
  Wei}}]{wei2006}
\bibinfo{author}{\bibfnamefont{Y.-H.} \bibnamefont{Li}},
  \bibinfo{author}{\bibfnamefont{X.~G.} \bibnamefont{Gong}}, \bibnamefont{and}
  \bibinfo{author}{\bibfnamefont{S.-H.} \bibnamefont{Wei}},
  \bibinfo{journal}{Phys. Rev. B} \textbf{\bibinfo{volume}{73}},
  \bibinfo{pages}{245206} (\bibinfo{year}{2006}).

\bibitem[{\citenamefont{Li et~al.}(2009)\citenamefont{Li, Walsh, Chen, Yin,
  Yang, Li, Da~Silva, Gong, and Wei}}]{wei2009}
\bibinfo{author}{\bibfnamefont{Y.-H.} \bibnamefont{Li}},
  \bibinfo{author}{\bibfnamefont{A.}~\bibnamefont{Walsh}},
  \bibinfo{author}{\bibfnamefont{S.}~\bibnamefont{Chen}},
  \bibinfo{author}{\bibfnamefont{W.-J.} \bibnamefont{Yin}},
  \bibinfo{author}{\bibfnamefont{J.-H.} \bibnamefont{Yang}},
  \bibinfo{author}{\bibfnamefont{J.}~\bibnamefont{Li}},
  \bibinfo{author}{\bibfnamefont{J.~L.~F.} \bibnamefont{Da~Silva}},
  \bibinfo{author}{\bibfnamefont{X.~G.} \bibnamefont{Gong}}, \bibnamefont{and}
  \bibinfo{author}{\bibfnamefont{S.-H.} \bibnamefont{Wei}},
  \bibinfo{journal}{Appl. Phys. Lett.} \textbf{\bibinfo{volume}{94}},
  \bibinfo{pages}{212109} (\bibinfo{year}{2009}).

\bibitem[{\citenamefont{Sj\"{o}stedt et~al.}(2000)\citenamefont{Sj\"{o}stedt,
  Nordstr\"{o}m, and Singh}}]{Sjostedt2000}
\bibinfo{author}{\bibfnamefont{E.}~\bibnamefont{Sj\"{o}stedt}},
  \bibinfo{author}{\bibfnamefont{L.}~\bibnamefont{Nordstr\"{o}m}},
  \bibnamefont{and} \bibinfo{author}{\bibfnamefont{D.~J.} \bibnamefont{Singh}},
  \bibinfo{journal}{Solid State Comm.} \textbf{\bibinfo{volume}{114}},
  \bibinfo{pages}{15} (\bibinfo{year}{2000}).

\bibitem[{\citenamefont{Madsen et~al.}(2001)\citenamefont{Madsen, Blaha,
  Schwarz, Sj\"{o}stedt, and Nordstr\"{o}m}}]{Madsen2001}
\bibinfo{author}{\bibfnamefont{G.~K.~H.} \bibnamefont{Madsen}},
  \bibinfo{author}{\bibfnamefont{P.}~\bibnamefont{Blaha}},
  \bibinfo{author}{\bibfnamefont{K.}~\bibnamefont{Schwarz}},
  \bibinfo{author}{\bibfnamefont{E.}~\bibnamefont{Sj\"{o}stedt}},
  \bibnamefont{and}
  \bibinfo{author}{\bibfnamefont{L.}~\bibnamefont{Nordstr\"{o}m}},
  \bibinfo{journal}{Phys. Rev. B} \textbf{\bibinfo{volume}{64}},
  \bibinfo{pages}{195134} (\bibinfo{year}{2001}).

\bibitem[{\citenamefont{Singh}(1994)}]{Singh1994}
\bibinfo{author}{\bibfnamefont{D.}~\bibnamefont{Singh}},
  \emph{\bibinfo{title}{Plane waves, pseudopotentials and the LAPW method}}
  (\bibinfo{publisher}{Kluwer Academic}, \bibinfo{year}{1994}).

\bibitem[{\citenamefont{Schwarz et~al.}(2001)\citenamefont{Schwarz, Blaha, and
  Madsen}}]{Schwarz2001}
\bibinfo{author}{\bibfnamefont{K.}~\bibnamefont{Schwarz}},
  \bibinfo{author}{\bibfnamefont{P.}~\bibnamefont{Blaha}}, \bibnamefont{and}
  \bibinfo{author}{\bibfnamefont{G.~K.~H.} \bibnamefont{Madsen}},
  \bibinfo{journal}{Comp. Phys. Commun.}  (\bibinfo{year}{2001}).

\bibitem[{\citenamefont{te~Velde}(1990)}]{BAND0}
\bibinfo{author}{\bibfnamefont{G.}~\bibnamefont{te~Velde}},
  \emph{\bibinfo{title}{Ph. \uppercase{D.} \uppercase{t}hesis,
  \uppercase{V}rije \uppercase{U}niversiteit, \uppercase{A}msterdam}}
  (\bibinfo{year}{1990}).

\bibitem[{\citenamefont{te~Velde and Baerends}(1991)}]{Velde1991}
\bibinfo{author}{\bibfnamefont{G.}~\bibnamefont{te~Velde}} \bibnamefont{and}
  \bibinfo{author}{\bibfnamefont{E.~J.} \bibnamefont{Baerends}},
  \bibinfo{journal}{Phys. Rev. B} \textbf{\bibinfo{volume}{44}},
  \bibinfo{pages}{7888} (\bibinfo{year}{1991}).

\bibitem[{\citenamefont{te~Velde et~al.}()\citenamefont{te~Velde, Baerends,
  Philipsen, Wiesenekker, Groeneveld, Berger, de~Boeij, Klooster, Kootstra,
  Romaniello et~al.}}]{BAND}
\bibinfo{author}{\bibfnamefont{G.}~\bibnamefont{te~Velde}},
  \bibinfo{author}{\bibfnamefont{E.~J.} \bibnamefont{Baerends}},
  \bibinfo{author}{\bibfnamefont{P.~H.~T.} \bibnamefont{Philipsen}},
  \bibinfo{author}{\bibfnamefont{G.}~\bibnamefont{Wiesenekker}},
  \bibinfo{author}{\bibfnamefont{J.~A.} \bibnamefont{Groeneveld}},
  \bibinfo{author}{\bibfnamefont{J.~A.} \bibnamefont{Berger}},
  \bibinfo{author}{\bibfnamefont{P.~L.} \bibnamefont{de~Boeij}},
  \bibinfo{author}{\bibfnamefont{R.}~\bibnamefont{Klooster}},
  \bibinfo{author}{\bibfnamefont{F.}~\bibnamefont{Kootstra}},
  \bibinfo{author}{\bibfnamefont{P.}~\bibnamefont{Romaniello}},
  \bibinfo{author}{\bibfnamefont{J. G.}~\bibnamefont{Snijders}},
  \bibinfo{author}{\bibfnamefont{E. S.}~\bibnamefont{Kadantsev}},
  \bibnamefont{and} \bibinfo{author}{\bibfnamefont{T.}~\bibnamefont{Ziegler}},
   \emph{\bibinfo{title}{Band 2008.01}},
  \bibinfo{note}{\uppercase{SCM}: Theoretical Chemistry, Vrije Universiteit,
  Amsterdam, The Netherlands}.

\bibitem[{\citenamefont{van Lenthe and Baerends}(1993)}]{vanLenthe1993}
\bibinfo{author}{\bibfnamefont{E.}~\bibnamefont{van Lenthe}} \bibnamefont{and}
  \bibinfo{author}{\bibfnamefont{J.~G.} \bibnamefont{Baerends},
  \bibfnamefont{E.~J.and~Snijders}}, \bibinfo{journal}{J. Chem. Phys.}
  \textbf{\bibinfo{volume}{99}}, \bibinfo{pages}{4597} (\bibinfo{year}{1993}).

\bibitem[{\citenamefont{van Lenthe et~al.}(1994)\citenamefont{van Lenthe,
  Baerends, and Snijders}}]{vanLenthe1994}
\bibinfo{author}{\bibfnamefont{E.}~\bibnamefont{van Lenthe}},
  \bibinfo{author}{\bibfnamefont{E.~J.} \bibnamefont{Baerends}},
  \bibnamefont{and} \bibinfo{author}{\bibfnamefont{E.~J.}
  \bibnamefont{Snijders}}, \bibinfo{journal}{J. Chem. Phys.}
  \textbf{\bibinfo{volume}{101}}, \bibinfo{pages}{9783} (\bibinfo{year}{1994}).

\bibitem[{\citenamefont{van Lenthe et~al.}(1996{\natexlab{a}})\citenamefont{van
  Lenthe, Snijders, and Baerends}}]{vanLenthe1996A}
\bibinfo{author}{\bibfnamefont{E.}~\bibnamefont{van Lenthe}},
  \bibinfo{author}{\bibfnamefont{J.~G.} \bibnamefont{Snijders}},
  \bibnamefont{and} \bibinfo{author}{\bibfnamefont{E.~J.}
  \bibnamefont{Baerends}}, \bibinfo{journal}{J. Chem. Phys.}
  \textbf{\bibinfo{volume}{105}}, \bibinfo{pages}{6505}
  (\bibinfo{year}{1996}{\natexlab{a}}).

\bibitem[{\citenamefont{van Lenthe et~al.}(1996{\natexlab{b}})\citenamefont{van
  Lenthe, van Leeuwen, Baerends, and Snijders}}]{vanLenthe1996B}
\bibinfo{author}{\bibfnamefont{E.}~\bibnamefont{van Lenthe}},
  \bibinfo{author}{\bibfnamefont{R.}~\bibnamefont{van Leeuwen}},
  \bibinfo{author}{\bibfnamefont{E.~J.} \bibnamefont{Baerends}},
  \bibnamefont{and} \bibinfo{author}{\bibfnamefont{J.~G.}
  \bibnamefont{Snijders}}, \bibinfo{journal}{Int. J. of Quant. Chem.}
  \textbf{\bibinfo{volume}{57}}, \bibinfo{pages}{281}
  (\bibinfo{year}{1996}{\natexlab{b}}).

\bibitem[{\citenamefont{van Lenthe et~al.}(1999)\citenamefont{van Lenthe,
  Ehlers, and Baerends}}]{vanLenthe1999}
\bibinfo{author}{\bibfnamefont{E.}~\bibnamefont{van Lenthe}},
  \bibinfo{author}{\bibfnamefont{A.~E.} \bibnamefont{Ehlers}},
  \bibnamefont{and} \bibinfo{author}{\bibfnamefont{E.~J.}
  \bibnamefont{Baerends}}, \bibinfo{journal}{J. Chem. Phys.}
  \textbf{\bibinfo{volume}{110}}, \bibinfo{pages}{8943} (\bibinfo{year}{1999}).

\bibitem[{\citenamefont{Kadantsev and Ziegler}(2009)}]{Kadantsev2009G}
\bibinfo{author}{\bibfnamefont{E.~S.} \bibnamefont{Kadantsev}}
  \bibnamefont{and} \bibinfo{author}{\bibfnamefont{T.}~\bibnamefont{Ziegler}},
  \bibinfo{journal}{J. Phys. Chem. A} \textbf{\bibinfo{volume}{113}},
  \bibinfo{pages}{1327} (\bibinfo{year}{2009}).

\bibitem[{\citenamefont{Philipsen et~al.}(1997)\citenamefont{Philipsen, van
  Lenthe, and Baerends}}]{Philipsen1997}
\bibinfo{author}{\bibfnamefont{P.~H.~T.} \bibnamefont{Philipsen}},
  \bibinfo{author}{\bibfnamefont{E.}~\bibnamefont{van Lenthe}},
  \bibnamefont{and} \bibinfo{author}{\bibfnamefont{E.~J.}
  \bibnamefont{Baerends}}, \bibinfo{journal}{Phys. Rev. B}
  \textbf{\bibinfo{volume}{56}}, \bibinfo{pages}{13556} (\bibinfo{year}{1997}).

\bibitem[{\citenamefont{Olsen et~al.}(2003)\citenamefont{Olsen, Philipsen, and
  Baerends}}]{Olsen2003}
\bibinfo{author}{\bibfnamefont{R.~A.} \bibnamefont{Olsen}},
  \bibinfo{author}{\bibfnamefont{P.~H.~T.} \bibnamefont{Philipsen}},
  \bibnamefont{and} \bibinfo{author}{\bibfnamefont{E.~J.}
  \bibnamefont{Baerends}}, \bibinfo{journal}{J. of Chem. Phys.}
  \textbf{\bibinfo{volume}{119}}, \bibinfo{pages}{4522} (\bibinfo{year}{2003}).

\bibitem[{\citenamefont{Romaniello and de~Boeij}(2005)}]{Romaniello2005}
\bibinfo{author}{\bibfnamefont{P.}~\bibnamefont{Romaniello}} \bibnamefont{and}
  \bibinfo{author}{\bibfnamefont{P.~L.} \bibnamefont{de~Boeij}},
  \bibinfo{journal}{J. Chem. Phys.} \textbf{\bibinfo{volume}{122}},
  \bibinfo{pages}{164303} (\bibinfo{year}{2005}).

\bibitem[{\citenamefont{Romaniello and de~Boeij}(2007)}]{Romaniello2007}
\bibinfo{author}{\bibfnamefont{P.}~\bibnamefont{Romaniello}} \bibnamefont{and}
  \bibinfo{author}{\bibfnamefont{P.}~\bibnamefont{de~Boeij}},
  \bibinfo{journal}{J. Chem. Phys.} \textbf{\bibinfo{volume}{127}},
  \bibinfo{pages}{174111} (\bibinfo{year}{2007}).

\bibitem[{\citenamefont{Kootstra et~al.}(2000)\citenamefont{Kootstra, de~Boeij,
  and Snijders}}]{Kootstra2000}
\bibinfo{author}{\bibfnamefont{F.}~\bibnamefont{Kootstra}},
  \bibinfo{author}{\bibfnamefont{P.~L.} \bibnamefont{de~Boeij}},
  \bibnamefont{and} \bibinfo{author}{\bibfnamefont{J.~G.}
  \bibnamefont{Snijders}}, \bibinfo{journal}{Phys. Rev. B}
  \textbf{\bibinfo{volume}{62}}, \bibinfo{pages}{7071} (\bibinfo{year}{2000}).

\bibitem[{\citenamefont{Hoffmann}(1988)}]{Hoffmann1988}
\bibinfo{author}{\bibnamefont{Hoffmann}}, \bibinfo{journal}{Rev. Mod. Phys.}
  \textbf{\bibinfo{volume}{60}}, \bibinfo{pages}{602} (\bibinfo{year}{1988}).

\bibitem[{EXC()}]{EXCITING}
\emph{\bibinfo{title}{http://exciting.sourceforge.net; version 0.9.151.}}

\bibitem[{\citenamefont{Boerrigter et~al.}(1988)\citenamefont{Boerrigter,
  te~Velde, and Baerends}}]{Boerrigter1988}
\bibinfo{author}{\bibfnamefont{P.~M.} \bibnamefont{Boerrigter}},
  \bibinfo{author}{\bibfnamefont{G.}~\bibnamefont{te~Velde}}, \bibnamefont{and}
  \bibinfo{author}{\bibfnamefont{E.~J.} \bibnamefont{Baerends}},
  \bibinfo{journal}{Int. J. Quantum Chem.} \textbf{\bibinfo{volume}{33}},
  \bibinfo{pages}{87} (\bibinfo{year}{1988}).

\bibitem[{\citenamefont{Wiesenekker and Baerends}(1991)}]{Wiesenekker1991}
\bibinfo{author}{\bibfnamefont{G.}~\bibnamefont{Wiesenekker}} \bibnamefont{and}
  \bibinfo{author}{\bibfnamefont{E.~J.} \bibnamefont{Baerends}},
  \bibinfo{journal}{J. of Phys.: Condensed Matter}
  \textbf{\bibinfo{volume}{3}}, \bibinfo{pages}{6721} (\bibinfo{year}{1991}).

\bibitem[{\citenamefont{Wiesenekker et~al.}(1988)\citenamefont{Wiesenekker,
  te~Velde, and Baerends}}]{Wiesenekker1988}
\bibinfo{author}{\bibfnamefont{G.}~\bibnamefont{Wiesenekker}},
  \bibinfo{author}{\bibfnamefont{G.}~\bibnamefont{te~Velde}}, \bibnamefont{and}
  \bibinfo{author}{\bibfnamefont{E.~J.} \bibnamefont{Baerends}},
  \bibinfo{journal}{J. of Phys.: Condensed Matter}
  \textbf{\bibinfo{volume}{21}}, \bibinfo{pages}{4263} (\bibinfo{year}{1988}).

\bibitem[{\citenamefont{Vosko et~al.}(1980)\citenamefont{Vosko, Wilk, and
  Nusair}}]{Vosko1980}
\bibinfo{author}{\bibfnamefont{S.~H.} \bibnamefont{Vosko}},
  \bibinfo{author}{\bibfnamefont{L.}~\bibnamefont{Wilk}}, \bibnamefont{and}
  \bibinfo{author}{\bibfnamefont{M.}~\bibnamefont{Nusair}},
  \bibinfo{journal}{Can. J. Phys.} \textbf{\bibinfo{volume}{58}},
  \bibinfo{pages}{1200} (\bibinfo{year}{1980}).

\bibitem[{\citenamefont{Perdew and Wang}(1992)}]{Perdew1992}
\bibinfo{author}{\bibfnamefont{J.~P.} \bibnamefont{Perdew}} \bibnamefont{and}
  \bibinfo{author}{\bibfnamefont{Y.}~\bibnamefont{Wang}},
  \bibinfo{journal}{Phys. Rev. B} \textbf{\bibinfo{volume}{45}}
  (\bibinfo{year}{1992}).

\bibitem[{\citenamefont{Murnaghan}(1937)}]{Murnaghan1937}
\bibinfo{author}{\bibfnamefont{F.~D.} \bibnamefont{Murnaghan}},
  \bibinfo{journal}{Am. J. Math.} \textbf{\bibinfo{volume}{49}}
  (\bibinfo{year}{1937}).

\bibitem[{\citenamefont{Mulliken}(1955)}]{Mulliken1955}
\bibinfo{author}{\bibfnamefont{R.~S.} \bibnamefont{Mulliken}},
  \bibinfo{journal}{J. Chem. Phys.} \textbf{\bibinfo{volume}{23}},
  \bibinfo{pages}{1833} (\bibinfo{year}{1955}).

\bibitem[{\citenamefont{Bickelhaupt et~al.}(1996)\citenamefont{Bickelhaupt, van
  Eikema~Hommes, Fonseca~Guerra, and Baerends}}]{Bickelhaupt1996}
\bibinfo{author}{\bibfnamefont{F.~M.} \bibnamefont{Bickelhaupt}},
  \bibinfo{author}{\bibfnamefont{N.~J.~R.} \bibnamefont{van Eikema~Hommes}},
  \bibinfo{author}{\bibfnamefont{C.}~\bibnamefont{Fonseca~Guerra}},
  \bibnamefont{and} \bibinfo{author}{\bibfnamefont{E.~J.}
  \bibnamefont{Baerends}}, \bibinfo{journal}{Organometallics}
  \textbf{\bibinfo{volume}{15}}, \bibinfo{pages}{2923} (\bibinfo{year}{1996}).

\bibitem[{\citenamefont{van~den Hoek et~al.}(1989)\citenamefont{van~den Hoek,
  Baerends, and van Santen}}]{Hoek1989}
\bibinfo{author}{\bibfnamefont{P.}~\bibnamefont{van~den Hoek}},
  \bibinfo{author}{\bibfnamefont{E.}~\bibnamefont{Baerends}}, \bibnamefont{and}
  \bibinfo{author}{\bibfnamefont{R.}~\bibnamefont{van Santen}},
  \bibinfo{journal}{J. Phys. Chem.} \textbf{\bibinfo{volume}{93}},
  \bibinfo{pages}{6469} (\bibinfo{year}{1989}).

\bibitem[{\citenamefont{Madelung}(1987)}]{LB}
\bibinfo{editor}{\bibfnamefont{O.}~\bibnamefont{Madelung}}, ed.,
  \emph{\bibinfo{title}{Landolt-B\"{o}rnstein, New Series}}, vol.
  \bibinfo{volume}{22a} (\bibinfo{publisher}{Springer-Verlag},
  \bibinfo{address}{Berlin}, \bibinfo{year}{1987}).

\bibitem[{\citenamefont{Nelmes and McMahon}(1998)}]{Nelmes1998}
\bibinfo{author}{\bibfnamefont{R.~J.} \bibnamefont{Nelmes}} \bibnamefont{and}
  \bibinfo{author}{\bibfnamefont{M.~I.} \bibnamefont{McMahon}},
  \bibinfo{journal}{Semicond. Semimetals} \textbf{\bibinfo{volume}{54}},
  \bibinfo{pages}{145} (\bibinfo{year}{1998}).

\bibitem[{\citenamefont{McMahon and Nelmes}(1997)}]{McMahon1997}
\bibinfo{author}{\bibfnamefont{M.~I.} \bibnamefont{McMahon}} \bibnamefont{and}
  \bibinfo{author}{\bibfnamefont{R.~J.} \bibnamefont{Nelmes}},
  \bibinfo{journal}{Phys. Rev. Lett.} \textbf{\bibinfo{volume}{78}},
  \bibinfo{pages}{3697} (\bibinfo{year}{1997}).

\bibitem[{\citenamefont{Vohra et~al.}(1985)\citenamefont{Vohra, Weir, and
  Ruoff}}]{Vohra1985}
\bibinfo{author}{\bibfnamefont{Y.~G.} \bibnamefont{Vohra}},
  \bibinfo{author}{\bibfnamefont{S.~T.} \bibnamefont{Weir}}, \bibnamefont{and}
  \bibinfo{author}{\bibfnamefont{A.~L.} \bibnamefont{Ruoff}},
  \bibinfo{journal}{Phys. Rev. B} \textbf{\bibinfo{volume}{31}},
  \bibinfo{pages}{7344} (\bibinfo{year}{1985}).

\bibitem[{\citenamefont{Menoni and Spain}(1987)}]{Menoni1987}
\bibinfo{author}{\bibfnamefont{C.~S.} \bibnamefont{Menoni}} \bibnamefont{and}
  \bibinfo{author}{\bibfnamefont{I.~L.} \bibnamefont{Spain}},
  \bibinfo{journal}{Phys. Rev. B} \textbf{\bibinfo{volume}{35}},
  \bibinfo{pages}{7520} (\bibinfo{year}{1987}).

\bibitem[{\citenamefont{Nelmes et~al.}(1995)\citenamefont{Nelmes, McMahon,
  Wright, Allan, Liu, and Loveday}}]{Nelmes1995}
\bibinfo{author}{\bibfnamefont{R.~J.} \bibnamefont{Nelmes}},
  \bibinfo{author}{\bibfnamefont{M.~I.} \bibnamefont{McMahon}},
  \bibinfo{author}{\bibfnamefont{N.~G.} \bibnamefont{Wright}},
  \bibinfo{author}{\bibfnamefont{D.~R.} \bibnamefont{Allan}},
  \bibinfo{author}{\bibfnamefont{H.}~\bibnamefont{Liu}}, \bibnamefont{and}
  \bibinfo{author}{\bibfnamefont{J.~S.} \bibnamefont{Loveday}},
  \bibinfo{journal}{J. Phys. Chem. Solids} \textbf{\bibinfo{volume}{56}},
  \bibinfo{pages}{539} (\bibinfo{year}{1995}).

\bibitem[{\citenamefont{Christensen}(1986)}]{Christensen1986}
\bibinfo{author}{\bibfnamefont{N.~E.} \bibnamefont{Christensen}},
  \bibinfo{journal}{Phys. Rev. B} \textbf{\bibinfo{volume}{33}},
  \bibinfo{pages}{5096} (\bibinfo{year}{1986}).

\bibitem[{\citenamefont{Zhang and Cohen}(1987)}]{Zhang1987}
\bibinfo{author}{\bibfnamefont{S.~B.} \bibnamefont{Zhang}} \bibnamefont{and}
  \bibinfo{author}{\bibfnamefont{M.~L.} \bibnamefont{Cohen}},
  \bibinfo{journal}{Phys. Rev. B} \textbf{\bibinfo{volume}{35}},
  \bibinfo{pages}{7604} (\bibinfo{year}{1987}).

\bibitem[{\citenamefont{Csonka et~al.}(2009)\citenamefont{Csonka, Perdew,
  Ruzsinszky, Philipsen, Leb\`{e}gue, Paier, Vydrov, and
  \'{A}ngy\'{a}n}}]{Perdew2009}
\bibinfo{author}{\bibfnamefont{G.~I.} \bibnamefont{Csonka}},
  \bibinfo{author}{\bibfnamefont{J.~P.} \bibnamefont{Perdew}},
  \bibinfo{author}{\bibfnamefont{A.}~\bibnamefont{Ruzsinszky}},
  \bibinfo{author}{\bibfnamefont{P.~H.~T.} \bibnamefont{Philipsen}},
  \bibinfo{author}{\bibfnamefont{S.}~\bibnamefont{Leb\`{e}gue}},
  \bibinfo{author}{\bibfnamefont{J.}~\bibnamefont{Paier}},
  \bibinfo{author}{\bibfnamefont{O.~A.} \bibnamefont{Vydrov}},
  \bibnamefont{and} \bibinfo{author}{\bibfnamefont{J.~G.}
  \bibnamefont{\'{A}ngy\'{a}n}}, \bibinfo{journal}{Phys. Rev. B}
  \textbf{\bibinfo{volume}{79}}, \bibinfo{pages}{155107}
  (\bibinfo{year}{2009}).

\bibitem[{\citenamefont{von Barth}(2004)}]{vonBarth2004}
\bibinfo{author}{\bibfnamefont{U.}~\bibnamefont{von Barth}},
  \bibinfo{journal}{Physica Scripta} \textbf{\bibinfo{volume}{T109}},
  \bibinfo{pages}{9} (\bibinfo{year}{2004}).

\bibitem[{\citenamefont{Ziegler et~al.}(1981)\citenamefont{Ziegler, Snijders,
  and Baerends}}]{Ziegler1981}
\bibinfo{author}{\bibfnamefont{T.}~\bibnamefont{Ziegler}},
  \bibinfo{author}{\bibfnamefont{J.~G.} \bibnamefont{Snijders}},
  \bibnamefont{and} \bibinfo{author}{\bibfnamefont{E.~J.}
  \bibnamefont{Baerends}}, \bibinfo{journal}{J. Chem. Phys.}
  \textbf{\bibinfo{volume}{74}}, \bibinfo{pages}{1271} (\bibinfo{year}{1981}).

\bibitem[{\citenamefont{Hirshfeld}(1977)}]{Hirshfeld}
\bibinfo{author}{\bibfnamefont{F.~L.} \bibnamefont{Hirshfeld}},
  \bibinfo{journal}{Theor. Chim. Acta} \textbf{\bibinfo{volume}{44}},
  \bibinfo{pages}{129} (\bibinfo{year}{1977}).

\bibitem[{\citenamefont{Kadantsev and Schmider}(2008)}]{Kadantsev2008}
\bibinfo{author}{\bibfnamefont{E.}~\bibnamefont{Kadantsev}} \bibnamefont{and}
  \bibinfo{author}{\bibfnamefont{H.}~\bibnamefont{Schmider}},
  \bibinfo{journal}{Int. J. Quant. Chem.} \textbf{\bibinfo{volume}{108}},
  \bibinfo{pages}{1} (\bibinfo{year}{2008}).

\bibitem[{\citenamefont{Pyykko}(1988)}]{Pyykko1988}
\bibinfo{author}{\bibfnamefont{P.}~\bibnamefont{Pyykko}},
  \bibinfo{journal}{Chem. Rev.} \textbf{\bibinfo{volume}{88}},
  \bibinfo{pages}{563} (\bibinfo{year}{1988}).

\end{thebibliography}

\newpage

\begin{table}[h]
\caption{\label{TAB:IIIVS} 
\\Structural parameters of {\rm InP}, {\rm GaAs} and {\rm InAs} 
IIIA-VA semiconductors (zinc-blende crystal structure). 
LSDA equilibrium lattice constant $a_{eq}$ (\AA), 
bulk moduli $B$ (kbar=0.1GPa) 
and pressure derivatives $B'$. Experimental values are taken from Ref.~\cite{LB}.
The experimental bulk modulus is computed from the elastic constants $c_{11}$ and $c_{12}$ as $B=(c_{11}+2 c_{12})/3$.}
\begin{tabular*}{1.0\textwidth}{@{\extracolsep{\fill}}lccc}
\hline
\hline\\
    \multicolumn{1}{c}{Method}    
 &  \multicolumn{1}{c}{$a_{eq}$ (\AA)}
 &  \multicolumn{1}{c}{$B$ (kbar)}
 &  \multicolumn{1}{c}{$B'$}
 \\
\multicolumn{4}{c}{{\rm InP}}\\   
\hline
 ZORA FREL         & 5.831  & 699   &  4.5 \\ 
 ZORA SREL         & 5.833  & 697   &  4.5 \\
 NREL              & 5.873  & 709   &  4.4 \\
 APW+lo           & 5.823  & 685   &  4.4 \\  
 EXP.       & 5.869$^a$  & 710$^d$   &   \\
\multicolumn{4}{c}{{\rm GaAs}}\\   
\hline
ZORA FREL        &  5.607 &  736 &  4.4 \\
ZORA SREL       &  5.610 &  732 &  4.4 \\ 
NREL            &  5.627 &  745 &  4.5 \\ 
APW+lo          &  5.610 &  691 &  4.4 \\  
EXP.     &  5.653$^b$ &  756$^d$ &      \\
\multicolumn{4}{c}{{\rm InAs}}\\   
\hline
ZORA FREL        &  6.026  & 590  &  4.7 \\
ZORA SREL        &  6.029  & 590  &  4.8 \\ 
NREL            &  6.069  & 614  &  4.4 \\ 
APW+lo          &  6.033  & 511  &  4.6 \\  
EXP.      &  6.058$^c$  & 579$^d$  &      \\
\hline 
\multicolumn{4}{l}{Experimental temperature: $^a$T=291.15 K, $^b$T=298.15 K, $^c$T=300 K, $^d$room
temperature}\\
\end{tabular*}
\end{table}

\begin{table*}[h]
\caption{\label{TAB:IIIVE} 
\\Parameters of electronic structure for {\rm InP}, {\rm GaAs},
and {\rm InAs} IIIA-VA semiconductors 
(zinc-blende crystal structure). All energies are in eV. $E_{gap}$ -- energy gaps 
for transition between high-symmetry points of the Brillouin zone, 
$\Delta_{SO}^{\Gamma}$ and $\Delta_{SO}^{L}$ -- magnitude of spin-orbital splitting at the top of the 
valence band at ${\Gamma}$ and $L$ points, 
$E_d^{\Gamma}$ -- $d$ band position at the $\Gamma$ point, 
$\delta E_d^{\Gamma}$ -- $d$ band width at the $\Gamma$ point, and $UVBW$ - the upper valence band width.}
\begin{tabular*}{1.0\textwidth}{@{\extracolsep{\fill}}lcccccccc}
\hline
\hline\\
    \multicolumn{1}{c}{Method}    
 &  \multicolumn{1}{c}{$E_{gap}^{\Gamma_V \rightarrow \Gamma_C}$}
 &  \multicolumn{1}{c}{$E_{gap}^{\Gamma_V \rightarrow X_C}$}
 &  \multicolumn{1}{c}{$E_{gap}^{\Gamma_V \rightarrow L_C}$}
 &  \multicolumn{1}{c}{$\Delta_{SO}^{\Gamma}$}
 &  \multicolumn{1}{c}{$\Delta_{SO}^{L}$}
 &  \multicolumn{1}{c}{$E_d^{\Gamma}$}
 &  \multicolumn{1}{c}{$\delta E_d^{\Gamma}$}
 &  \multicolumn{1}{c}{UVBW} 
 \\
\multicolumn{9}{c}{{\rm InP}}\\   
\hline
 ZORA FREL     & 0.437  &  1.606  & 1.292  &  0.108   &  0.120 & -14.19 & 0.89 & 5.89 \\ 
 ZORA SREL    & 0.475  &  1.644  & 1.330  &  0.0     &  0.0   & -14.09 & 0.16 & 5.85 \\
 NREL         & 1.014  &  1.590  & 1.781  &  0.0     &  0.0   & -15.07 & 0.14 & 5.51 \\
 APW+lo       & 0.416  &  1.575  & 1.257  &  0.100   &  0.112 & -14.21 & 0.89 & 5.90 \\  
 EXP.         & 1.420  &  2.384  & 2.014  &  0.110   &  0.130 &        &      &        \\
\hline     
\multicolumn{9}{c}{{\rm GaAs}}\\   
 ZORA FREL     & 0.196  & 1.257 &  0.748  &  0.349   & 0.213 & -15.01 &  0.48 &  7.00 \\
 ZORA SREL      & 0.313  & 1.372 &  0.865  &  0.0     & 0.0   & -14.85 &  0.08 &  6.89 \\ 
 NREL           & 0.965  & 1.418 &  1.197  &  0.0     & 0.0   & -15.38 &  0.08 &  6.69 \\ 
 APW+lo       & 0.173  & 1.240 &  0.735  &  0.341   & 0.209 & -15.02 &  0.48 &  6.99 \\  
 EXP.         & 1.520  & 1.979 &  1.819  &  0.340   & 0.220 &        &       &       \\
\hline	     
\multicolumn{9}{c}{{\rm InAs}} \\  
ZORA FREL     & -0.577  & 1.325   & 0.661  &  0.363 & 0.262 &  -14.46  &  0.88  &  6.02 \\
ZORA SREL     & -0.459  & 1.444   & 0.784  &  0.0   & 0.0   &  -14.27  &  0.13  &  5.90 \\ 
NREL          &  0.405  & 1.453   & 1.337  &  0.0   & 0.0   &  -15.26  &  0.11  &  5.49 \\ 
APW+lo       & -0.594  & 1.325   & 0.653  &  0.351 & 0.250 &  -14.48  &  0.88  &  6.03 \\  
EXP.         &  0.420  & 2.24    & 1.133  &  0.380 & 0.267 &          &          &         \\
\hline  \\ 
\end{tabular*}
\end{table*}

\begin{table*}[h]
\caption{\label{TAB:DFTRES} 
\\Orbital populations of the band edges in {\rm InP}, {\rm GaAs}, and {\rm InAs} semiconductors 
(zinc-blende crystal structure) in  high-symmetry points of the Brillouin zone at the equilibrium 
lattice constants. SO -- split-off band, LH -- light hole band, HH -- heavy hole band, and CB -- 
conduction band minimum. ZORA FREL LSDA calculation.}
\begin{tabular*}{1.0\textwidth}{@{\extracolsep{\fill}}lcccccccccccc}
\hline
\hline\\
\multicolumn{13}{c}{{\rm InP}} \\  
\hline
    \multicolumn{1}{c}{}    
 &  \multicolumn{4}{c}{$\Gamma$}
 &  \multicolumn{4}{c}{$X$}
 &  \multicolumn{4}{c}{$L$}\\
    \multicolumn{1}{c}{}    
 &  \multicolumn{1}{c}{$In.s$}
 &  \multicolumn{1}{c}{$In.p$}
 &  \multicolumn{1}{c}{$P.s$}
 &  \multicolumn{1}{c}{$P.p$}
 &  \multicolumn{1}{c}{$In.s$}
 &  \multicolumn{1}{c}{$In.p$}
 &  \multicolumn{1}{c}{$P.s$}
 &  \multicolumn{1}{c}{$P.p$}
 &  \multicolumn{1}{c}{$In.s$}
 &  \multicolumn{1}{c}{$In.p$}
 &  \multicolumn{1}{c}{$P.s$}
 &  \multicolumn{1}{c}{$P.p$}\\
SO  & 0.000 & 0.112 & 0.000 & 0.809 & 0.619 & 0.000 & 0.000 &  0.327 & 0.451 & 0.328 & 0.042 & 0.179 \\
LH  & 0.000 & 0.072 & 0.000 & 0.878 & 0.000 & 0.360 & 0.000 &  0.630 & 0.000 & 0.237 & 0.000 & 0.752 \\
HH  & 0.000 & 0.072 & 0.000 & 0.878 & 0.000 & 0.339 & 0.000 &  0.645 & 0.000 & 0.209 & 0.000 & 0.786 \\
CB  & 0.920 & 0.000 & 0.067 & 0.000 & 0.000 & 0.621 & 0.030 &  0.000 & 0.430 & 0.252 & 0.014 & 0.234 \\
\multicolumn{13}{c}{{\rm GaAs}} \\  
\hline
    \multicolumn{1}{c}{}    
 &  \multicolumn{4}{c}{$\Gamma$}
 &  \multicolumn{4}{c}{$X$}
 &  \multicolumn{4}{c}{$L$}\\
    \multicolumn{1}{c}{}    
 &  \multicolumn{1}{c}{$Ga.s$}
 &  \multicolumn{1}{c}{$Ga.p$}
 &  \multicolumn{1}{c}{$As.s$}
 &  \multicolumn{1}{c}{$As.p$}
 &  \multicolumn{1}{c}{$Ga.s$}
 &  \multicolumn{1}{c}{$Ga.p$}
 &  \multicolumn{1}{c}{$As.s$}
 &  \multicolumn{1}{c}{$As.p$}
 &  \multicolumn{1}{c}{$Ga.s$}
 &  \multicolumn{1}{c}{$Ga.p$}
 &  \multicolumn{1}{c}{$As.s$}
 &  \multicolumn{1}{c}{$As.p$}\\
SO  & 0.000 &  0.118 & 0.000 & 0.755 & 0.523 & 0.000 & 0.000 & 0.402 & 0.452 &  0.352 & 0.015 & 0.183 \\
LH  & 0.000 &  0.128 & 0.000 & 0.738 & 0.000 & 0.435 & 0.000 & 0.539 & 0.000 &  0.294 & 0.000 & 0.661 \\
HH  & 0.000 &  0.128 & 0.000 & 0.738 & 0.000 &  0.453 & 0.000 & 0.524 &  0.000 &  0.279 & 0.000 & 0.636 \\
CB  & 0.728 &  0.000 & 0.216 & 0.000 & 0.000 &  0.556 & 0.040 & 0.000 & 0.305 &   0.298 &  0.092 & 0.130 \\
\multicolumn{13}{c}{{\rm InAs}} \\  
\hline
    \multicolumn{1}{c}{}    
 &  \multicolumn{4}{c}{$\Gamma$}
 &  \multicolumn{4}{c}{$X$}
 &  \multicolumn{4}{c}{$L$}\\
    \multicolumn{1}{c}{}    
 &  \multicolumn{1}{c}{$In.s$}
 &  \multicolumn{1}{c}{$In.p$}
 &  \multicolumn{1}{c}{$As.s$}
 &  \multicolumn{1}{c}{$As.p$}
 &  \multicolumn{1}{c}{$In.s$}
 &  \multicolumn{1}{c}{$In.p$}
 &  \multicolumn{1}{c}{$As.s$}
 &  \multicolumn{1}{c}{$As.p$}
 &  \multicolumn{1}{c}{$In.s$}
 &  \multicolumn{1}{c}{$In.p$}
 &  \multicolumn{1}{c}{$As.s$}
 &  \multicolumn{1}{c}{$As.p$}\\
SO  & 0.000 &  0.100 &  0.000 &  0.829 & 0.610 &  0.000 & 0.000 & 0.303 & 0.485 &  0.279 &  0.019 &  0.217 \\
LH  & 0.000 &  0.081 &  0.000 &  0.843 & 0.000 &  0.339 & 0.000 & 0.637 & 0.000 &   0.242 &  0.000 &  0.750 \\
HH  & 0.000 &  0.081 &  0.000 &  0.843 & 0.000 &  0.360 & 0.000 & 0.611 & 0.000 &   0.210 &  0.000 &  0.780 \\
CB  & 0.897 &  0.000 &  0.093 &  0.000 & 0.000 &  0.605 & 0.039 & 0.000 & 0.354 &   0.347 &  0.026 &  0.143 \\
\hline  \\ 
\end{tabular*}
\end{table*}

\begin{table}[h]
\caption{\label{TAB:IIIVDEFPOT} 
\\Relative volume deformation potentials $a_V$ (eV) for {\rm InP}, {\rm GaAs},
and {\rm InAs} IIIA-VA semiconductors (zinc-blende crystal structure) for specific transitions.
The different signs of the deformation potential is attributed to the 
``different nature" of the conduction band minimum.  Experimental volume deformation potentials 
are obtained from the direct band gap pressure dependence 
coefficient and bulk modulus.}
\begin{tabular*}{1.0\textwidth}{@{\extracolsep{\fill}}lccc}
\hline
\hline\\
    \multicolumn{1}{c}{Method}    
 &  \multicolumn{1}{c}{$a_V^{\Gamma_V \rightarrow \Gamma_C}$}
 &  \multicolumn{1}{c}{$a_V^{\Gamma_V \rightarrow X_C}$}
 &  \multicolumn{1}{c}{$a_V^{\Gamma_V \rightarrow L_C}$}
 \\
\multicolumn{4}{c}{{\rm InP}}\\   
\hline
 ZORA FREL      & -5.44 & 1.64  & -2.42  \\ 
 ZORA SREL       & -5.45 & 1.64  & -2.42  \\
 NREL            & -5.12 & 1.60  & -2.65  \\
 APW+lo        & -5.44 & 1.64  & -2.40  \\  
 EXP.          & -5.7  &      &        \\
\hline     
\multicolumn{4}{c}{{\rm GaAs}}\\   
 ZORA FREL     & -7.46 & 1.83 & -2.78  \\
 ZORA SREL      & -7.52 & 1.79 & -2.82  \\ 
 NREL           & -7.79 & 1.81 & -3.06  \\ 
 APW+lo       & -7.41 & 1.84 & -2.74  \\  
 EXP.         & -8.0,-9.2      &     &        \\
\hline	     
\multicolumn{4}{c}{{\rm InAs}} \\  
ZORA FREL     & -5.04 & 1.59 & -2.11  \\
ZORA SREL      & -5.09 & 1.56 & -2.15  \\ 
NREL           & -5.11 & 1.51 & -2.58  \\ 
APW+lo       & -5.01 & 1.59 & -2.09  \\  
EXP.         & -6.6  &      &        \\
\hline  \\ 
\end{tabular*}
\end{table}

\begin{table}[h]
\caption{\label{TAB:KPOINTS} 
\\Cartesian coordinates of special $\mathbf{k}$-points 
used to construct the band structure plots (special $\mathbf{k}$-points are shown with vertical lines).  
$a$ - lattice constant. }
\begin{tabular*}{1.0\textwidth}{@{\extracolsep{\fill}}ll}
\hline
\hline\\
    \multicolumn{1}{c}{Symbol} 
 &  \multicolumn{1}{c}{Coordinates}    
 \\
\hline
          & $\frac{2 \pi}{a} \left[1,-1/4,1/4 \right] $  \\
$L$       & $\frac{2 \pi}{a} \left[1/2,-1/2,1/2 \right] $  \\   
$\Gamma$  & $\frac{2 \pi}{a} \left[0,0,0 \right] $  \\   
          & $\frac{2 \pi}{a} \left[1,-1/4,1/4 \right] $  \\
$X$       & $\frac{2 \pi}{a} \left[1,0,0 \right] $  \\    
$\Gamma$  & $\frac{2 \pi}{a} \left[0,0,0 \right] $  \\   
$W$       & $\frac{2 \pi}{a} \left[1,0,1/2 \right] $  \\   
          & $\frac{2 \pi}{a} \left[1,-1/4,1/4 \right] $  \\
$\Gamma$  & $\frac{2 \pi}{a} \left[0,0,0 \right] $  \\
\hline  \\ 
\\
\end{tabular*}
\end{table}

\end{document}